\documentclass[3p]{elsarticle}

\usepackage{lineno,hyperref}
\usepackage{xspace}
\modulolinenumbers[1]
\usepackage{bm}
\usepackage{amssymb}
\usepackage{hyperref}
\usepackage{amsmath}
\usepackage{graphicx}
\usepackage{comment}
\usepackage{color}
\usepackage{float}

\newcommand{\s}{\ensuremath{\sqrt{s}}\xspace}
\newcommand{\gevc}{\ensuremath{\rm{GeV/}c}\xspace}
\newcommand{\pt}{\ensuremath{p_{\mathrm{T}}}\xspace}

\newcommand{\sccr}{SC-CR\xspace}
\newcommand{\blc}{BLC-CR~+~Ropes\xspace} 
\newcommand{\barionone}{\ensuremath{\Lambda_{c}^{+}/\rm{D}^0}\xspace} 
\newcommand{\bariontwo}{\ensuremath{\Sigma_{c}^{0,+,++}/\rm{D}^0}\xspace}
\newcommand{\mesonone}{\ensuremath{\rm{D}^{+}/\rm{D}^0}\xspace}
\newcommand{\mesontwo}{\ensuremath{\rm{D}^{*+}/\rm{D}^0}\xspace}
\newcommand{\zjet}{\ensuremath{z_{||}}\xspace}
\newcommand{\pth}{\ensuremath{p_{\mathrm{T}}^{\mathrm{H}}}\xspace}
\newcommand{\ptj}{\ensuremath{p_{\mathrm{T}}^{\mathrm{J}}}\xspace} 
\newcommand{\rt}{\ensuremath{R_{\mathrm{T}}}\xspace}

\journal{Journal of \LaTeX\ Templates}

\begin{document}

\begin{frontmatter}

\title{{\bf Impact of local and global event conditions in prompt charm hadron ratios}}

\author{Luis Carlos Cruz Peralta and Lizardo Valencia Palomo*}
\address{Departamento de Investigaci\'on en F\'isica, Universidad de Sonora, \\ Blvd. Luis Encinas y Rosales S/N, Col. Centro, Hermosillo, Sonora, M\'exico}

\cortext[mycorrespondingauthor]{lizardo.valencia@unison.mx}

\begin{abstract}

Fragmentation functions extracted from electron-proton and electron-positron collisions have shown not to be able to explain prompt charm hadron ratios in proton-proton collisions at the LHC energies. New models and event generator tunes have been developed to correctly describe the data from the ALICE experiment in proton-proton collisions at different center of mass energies. In this work Pythia 8.312 is used to implement two of those models: the \sccr and \blc. Both show a good performance to reproduce the data at \s = 13 TeV in different event conditions for the \barionone, \bariontwo, \mesonone and \mesontwo ratios. For the particular case when the prompt charm hadrons are within jets, all the ratios show modifications relative to the case where the presence of jets is not required. A study in two different ranges of the relative transverse activity is also conducted. In this case only the baryon to meson ratios show a dependence on the underlying event activity. Results indicate that baryon and meson hadronisation are not affected in the same way by local effects or global event conditions.

\end{abstract}


\end{frontmatter}


\section{Introduction} 

The theoretical framework used to study heavy hadron production relies on the factorization theorem from Quantum Chromodynamics (QCD) \cite{FactorizationTheo}. In this approach the cross section calculations for hadronic collisions are split into three elements: the Parton Distribution Functions (PDF), the partonic hard scattering and the hadronisation. Only the second step can be computed perturbatively in a power series of the strong running coupling constant. The hadronisation process assumes the universality of the fragmentation functions: results extracted from electron-proton (e$^-$p) and electron-positron (e$^-$e$^+$) collisions are still valid for proton-proton (pp) interactions \cite{UniversalityFF}. In this sense, measurements of the probability for a heavy quark to hadronise into a certain heavy hadron (fragmentation fractions) in different collision systems are crucial tests of the universality principle. Experimentally, this can be achieved by measuring the relative production rates of different heavy hadron species containing the same heavy quark.

For charm quark studies at the LHC, the ALICE experiment has reported baryon to meson, meson to meson and baryon to baryon ratios at \s = 13 TeV \cite{Multiplicity,Cascade13TeV,D0LambdaSigma,NonPrompt,Dmesons}. The fragmentation fractions extracted from these results point to the break down of the universality principle. This is, the probability of a charm quark to hadronise into a certain hadron is not the same in pp collisions as in e$^-$e$^+$ or e$^-$p collisions. For this reason, the data can not be reproduced by Pythia 8 event generator using only the standard Monash tune, as the fragmentation functions are parametrised using e$^-$p and e$^-$e$^+$ collisions \cite{Pythia8.3,MonashTune}. A much better description can be accomplished by models incorporating different hadronisation mechanisms. In Pythia 8, predictions are greatly improved if a new Color Reconnection (CR) approach, called Beyond Leading Color, is applied \cite{CR-BLC}. The Statistical Hadronisation Model plus Relativistic Quark Model assumes the existence of an augmented set of excited charm baryons that have not been detected yet and whose decay enhances the number of the ground state charm baryons \cite{SHM,RQM}. There are also theoretical frameworks where the hadronisation by fragmentation is complemented, or even replaced, by other mechanism. The POWLANG model assumes the formation of a deconfined state, similar to the one that is created in high energy heavy ion collisions, but at a much smaller scale where charm quarks undergo rescattering and hadron formation is completely driven by recombination \cite{POWLANG}. The Catania model also presupposes the presence of a fireball so that charm quark hadronisation arises as an interplay between fragmentation and coalescence with light quarks, that is indeed the dominant process for charm quarks with transverse momentum (\pt) close to zero \cite{Catania}. Finally, in the Quark re-Combination Model, as the name suggests, the hadronisation of charm quarks is entirely due to the re-combination with light quarks that are close in the velocity phase space \cite{QCM}. 



\section{The SC-CR and BLC-CR + Ropes models}

Pythia 8 is a Monte Carlo event generator commonly used in high energy physics to understand the underlying mechanisms of different particle collisions \cite{Pythia8.3}. In Pyhtia 8 a pp collision can be divided in three well defined components: the partonic interactions, the parton shower evolution and finally the hadronisation (fragmentation). The first stage contains the hard partonic scattering (involving the largest momentum transfer) and an ensemble of softer scatterings called Multiple Partonic Interactions (MPI). For heavy quarks ($Q$), production is implemented through gluon fussion $gg \rightarrow Q\overline{Q}$, light quark ($q$) and anti-quark anhilation $q\overline{q} \rightarrow Q\overline{Q}$, flavour excitation of $Q$ present in the PDF $Qg \rightarrow Qg$ and during the shower by gluon splitting $g \rightarrow Q\overline{Q}$. The products of the partonic scatterings are connected by color strings or fluxtubes, where the potential energy depends on the linear separation. As partons drift away, the strings break, giving rise to new $q\overline{q}$ pairs. 

Originally, the different MPI were treated independently but in order to properly describe pp data Pythia 8 introduces Color Reconnection (CR). In this framework partons from indenpendet MPI are color-connected following the string length minimization requirement, as this controls the number of hadrons generated during the string breaking. Further development of the CR has lead to the Beyond Leading Color (BLC-CR) approach that includes SU(3) color algebra rules \cite{CR-BLC}. This implies the possibility for color strings to create a junction when connecting three color dipoles or a system of a junction and anti-junction where each one is joining a pair of dipoles. The junctions present a novel baryon production method, besides the already existing diquark generation in the string fragmentation process. There are some available tunes based on this BLC-CR, where the main difference is related to the time between the formation of a string and the hadronisation. The BLC-CR mode 2, where all the QCD dipoles are causally connected, is cataloged as the default one. This model, with the addition of colour ropes (fusion of near by strings in transverse space) has been successfully applied to compute the total charm and anti-charm cross section by extrapolating LHC measurements to kinematic regions without existing data \cite{CharmModel}. 

There are current efforts to extend Pythia 8 to high energy heavy ion collisions by modeling the nuclei events as a superposition of $n$ nucleon-nucleon interactions \cite{Angantyr}. As the range of the strong force is around the size of the proton, a key element is to determine the interaction of partons from different nucleon collisions. For this reason a spatial constraint is introduced to restrict the separation among the color dipoles that are allowed to be color reconnected \cite{SCCRoriginal}. This characteristic gives birth to the name of the model: Spatially Constrained Color Reconnection (SC-CR). The SC-CR model is built on the BLC-CR but improves the junction hadronisation and introduces a re-tune of certain parameters. The tune is performed on pp collisions at \s = 7 TeV  using charged particle multiplicity and the mean transverse momentum. Among the effects of these changes is the modification of the baryon to meson ratios for light quarks. For heavy quark physics, further developments are applied to the junction formation \cite{SCCRcharm}. First, strings connected to a heavy quark have now a different length measure with respect to light quarks and massless gluons. Second, the junction system is hadronised in the rest frame of the charm or beauty quark. Finally, junction legs ordering is switched so the leg that contains a heavy quark is the first to be fragmented.   


\begin{figure*}
\begin{center}
\includegraphics[width=0.49\textwidth]{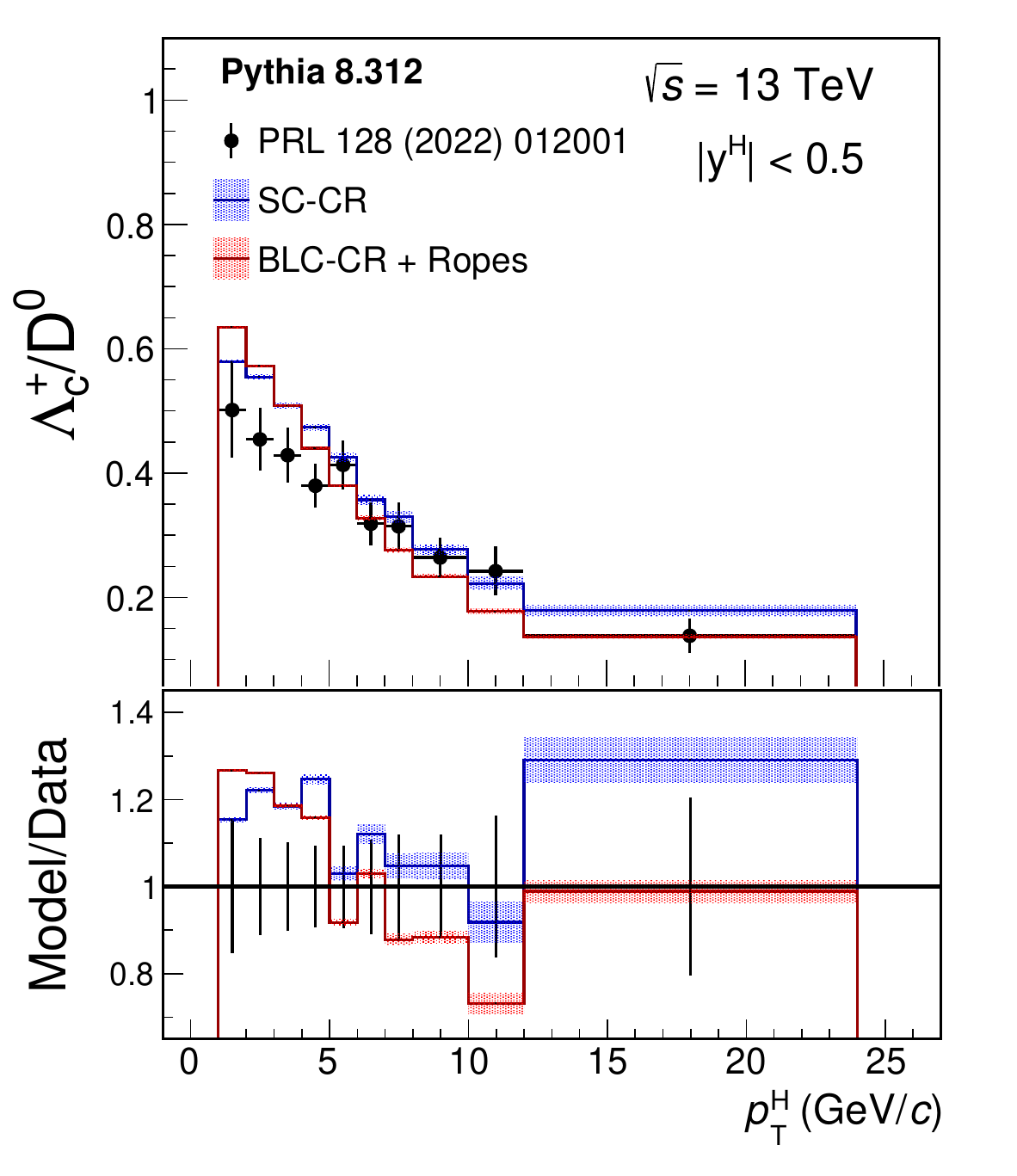}
\hspace{-0.4cm}
\includegraphics[width=0.49\textwidth]{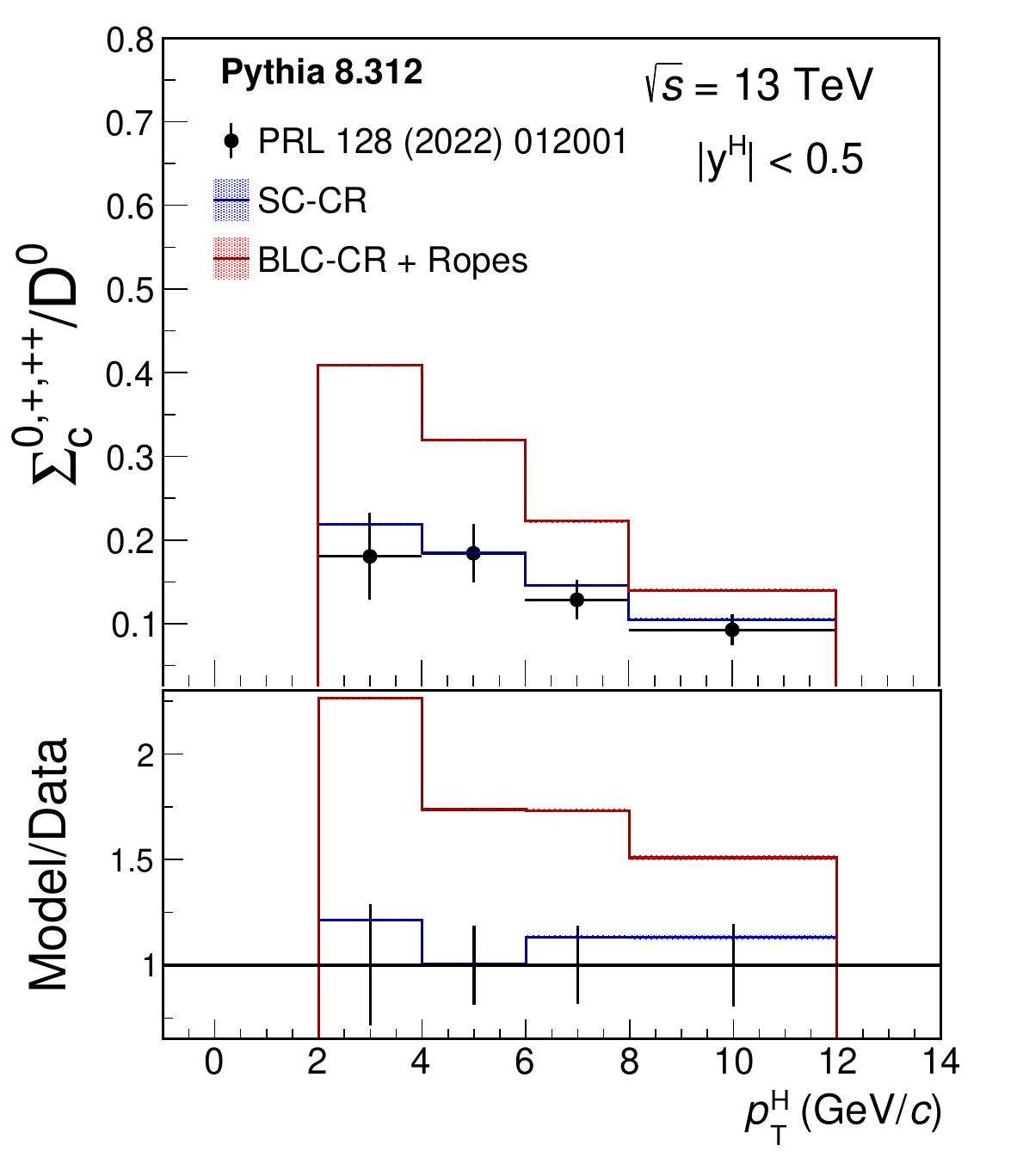}
\caption{Baryon to meson ratios as a function of \pth: \barionone (left) and \bariontwo (right). Black markers are the experimental data where the vertical bars are the quadratic sum of the statistical and systematic uncertainties while the horizontal bars indicate the bin width. Full lines represent the \sccr (blue) and \blc (red) predictions. Shaded areas around the lines are the statistical uncertainties from the simulation. Bottom panels show the model to data ratios.}
\label{DvsSbaryons}
\end{center}
\end{figure*}

\begin{figure*}
\begin{center}
\includegraphics[width=0.49\textwidth]{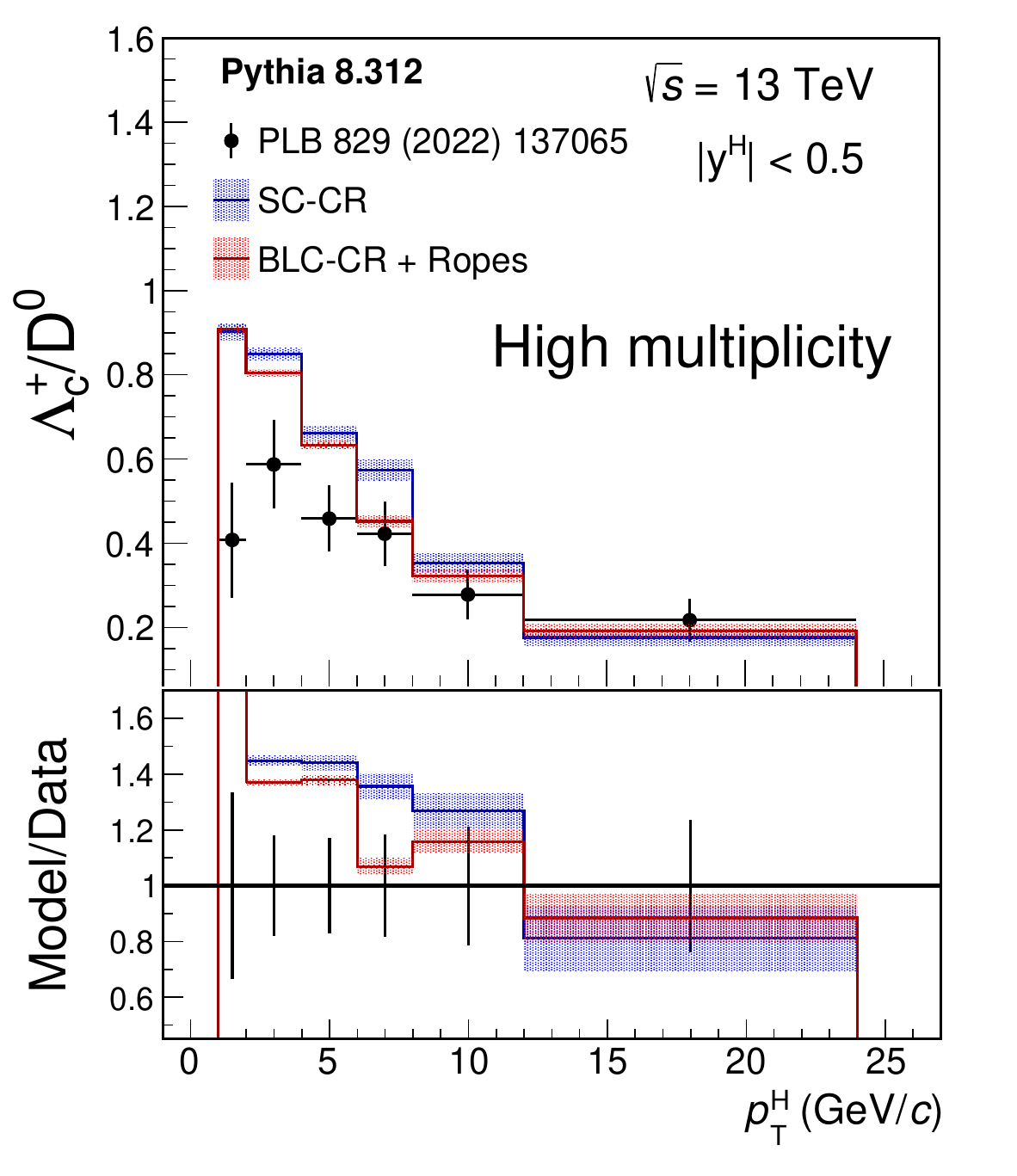}
\hspace{-0.4cm}
\includegraphics[width=0.49\textwidth]{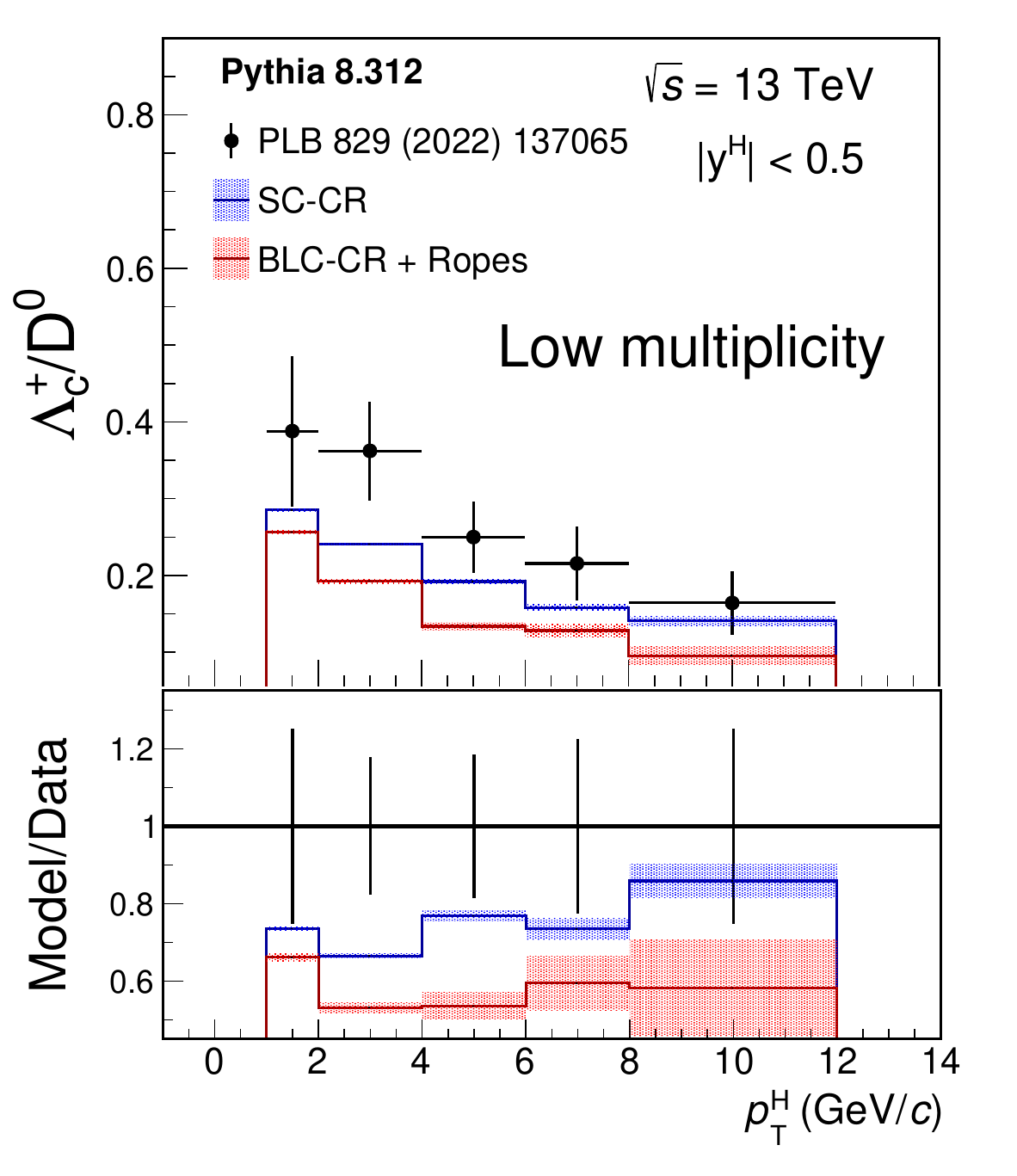}
\caption{\barionone ratios in two different multiplicity ranges as a function of \pth: high (left) and low (right) multiplicity. Black markers are the experimental data where the vertical bars are the quadratic sum of the statistical and systematic uncertainties while the horizontal bars indicate the bin width. Full lines represent the \sccr (blue) and \blc (red) predictions. Shaded areas around the lines are the statistical uncertainties from the simulation. Bottom panels show the model to data ratios.}
\label{DvsSmult}
\end{center}
\end{figure*}

\begin{figure*}
\begin{center}
\includegraphics[width=0.49\textwidth]{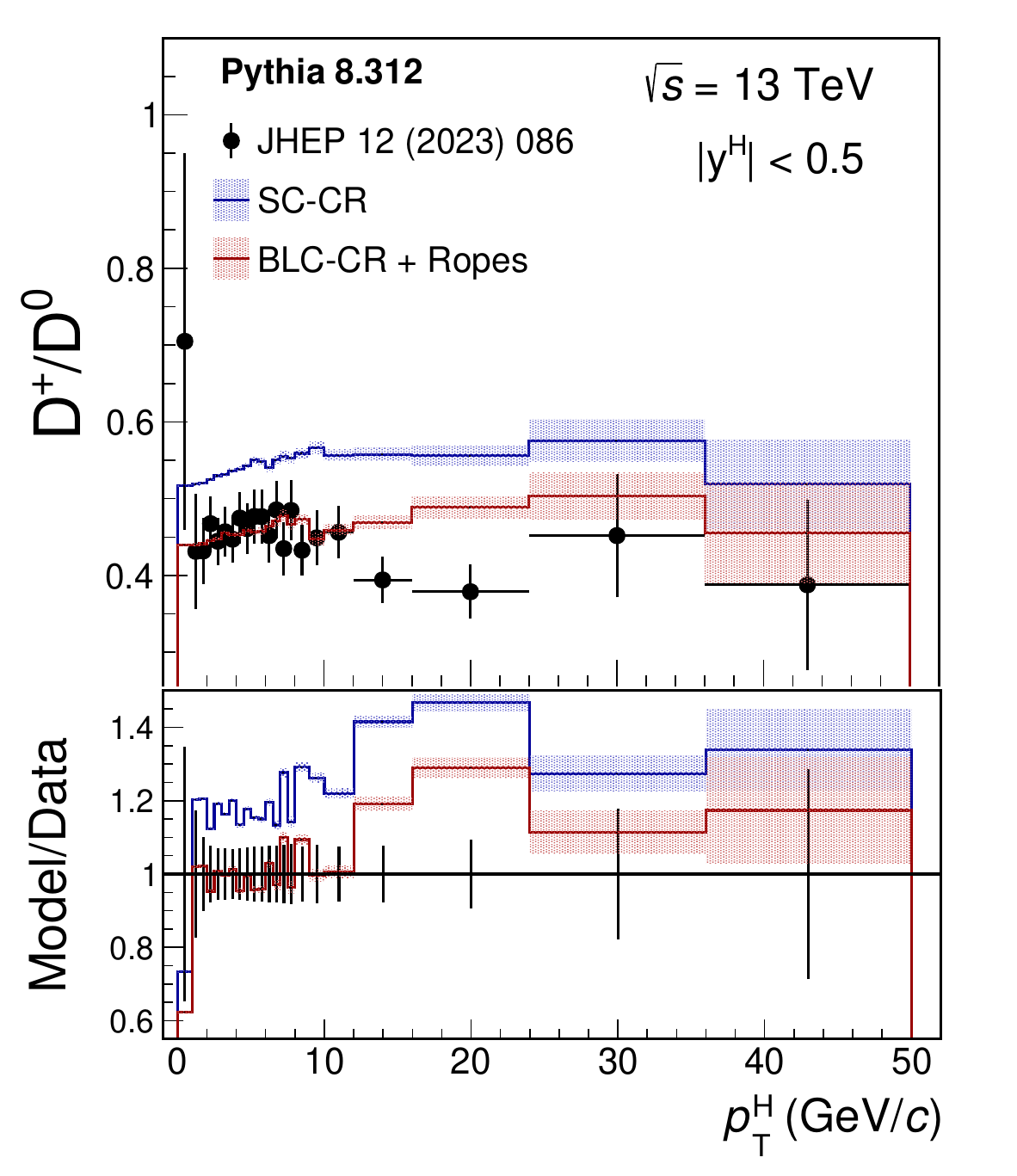}
\hspace{-0.4cm}
\includegraphics[width=0.49\textwidth]{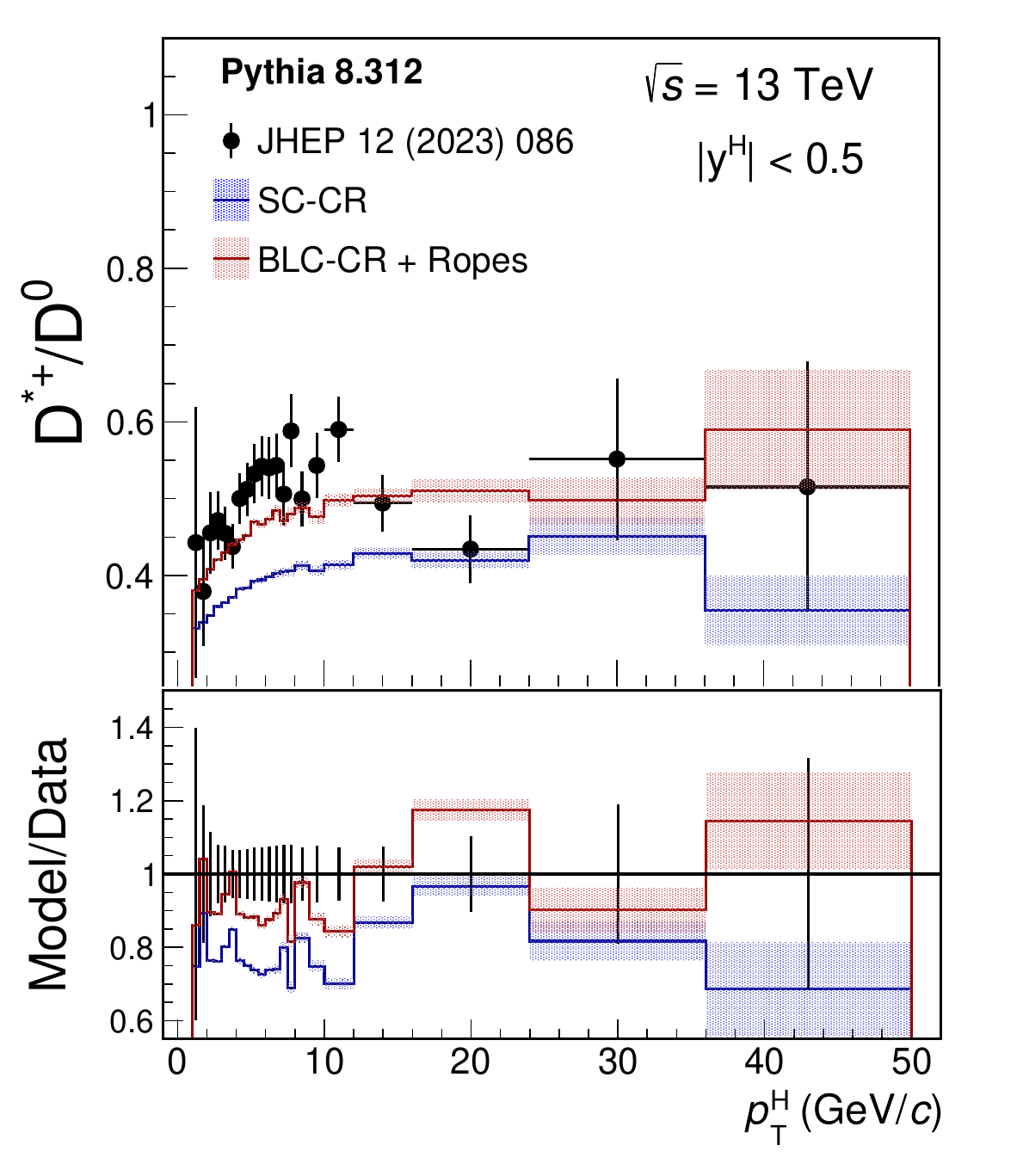}
\caption{Meson to meson ratios as a function of \pth: \mesonone (left) and \mesontwo (right). Black markers are the experimental data where the vertical bars are the quadratic sum of the statistical and systematic uncertainties while the horizontal bars indicate the bin width. Full lines represent the \sccr (blue) and \blc (red) predictions. Shaded areas around the lines are the statistical uncertainties from the simulation. Bottom panels show the model to data ratios.}
\label{DvsSmesons}
\end{center}
\end{figure*}


In this work the \sccr and the \blc models are used to study the behavior of prompt charm baryon to meson and meson to meson ratios in different event conditions. To this end, the present manuscript will focus on $\Lambda_c^+$/D$^0$, $\Sigma_c^{0,+,++}$/D$^0$, D$^+$/D$^0$ and D$^{*+}$/D$^0$ ratios measured at central rapidity ($|y|< 0.5$) in pp collisions at $\s$ = 13 TeV as reported by the ALICE experiment \cite{Multiplicity,D0LambdaSigma,Dmesons}. In figures \ref{DvsSbaryons} to \ref{DvsSmesons} the black markers are the experimental data where the vertical bars are the quadratic sum of the statistical and systematic uncertainties while the horizontal bars indicate the bin width. Notice that the \pt range is not the same in all plots. Pythia 8.312 is used to generate the \sccr (blue line) and \blc (red line) predictions. Shaded areas around the lines are the statistical uncertainties from the simulation. Bottom panels show the model to data ratios, indicating the ability of the model to reproduce the data. In these quotients, the vertical bars in the unit line are the errors from the data. 

Figure \ref{DvsSbaryons} and \ref{DvsSmult} show baryon to meson ratios. The former presents \barionone and \bariontwo, while the latter depicts only \barionone in two intervals that depend on the number of charged particles produced in the event. The multiplicity is determined based on the number of primary charged particles ($N_{\mathrm{ch}}$) measured in $|\eta|< 1$, where the decay products of the charm hadrons are not included into the multiplicity estimation to avoid any potential bias \cite{PrimaryChPart}. Events with $1 < N_{\mathrm{ch}} < 9$ are cataloged as low multiplicity, while those with $60 < N_{\mathrm{ch}} < 99$ are in the high multiplicity regime. In figure \ref{DvsSbaryons} only \blc overestimates the \bariontwo ratios. Figure \ref{DvsSmesons} contains meson to meson ratios (\mesonone and \mesontwo), where \blc has a great performance. 





\section{Charm hadrons in jets}

Further constraints to the models can be reached by studying the fragmentation of jets containing a heavy hadron, as these processes can be sensitive to the gluon splitting during the partonic shower. This has already been observed for the J/$\psi$, where CMS and LHCb measured its production in jets using the \pt ratio between the J/$\psi$ and the jet, a variable called $z$ \cite{JpsiInJetsCMS,JpsiInJetsLHCb}. For both experiments Pythia 8 over predicts the measurements when $z \approx 1$. This was a clear indication that another J/$\psi$ production mechanism was not being taken into account by the event generator. For this reason a new quarkonia shower, including J/$\psi$ production from quark and gluon splittings, was added to Pythia 8 \cite{OniaShower}. With this modification the description of the data is greatly improved \cite{JpsiUpsilonJets}.

Current charm hadron production in jets reported by ALICE include $\Lambda_c^+$ and D$^0$ \cite{D0Jets,LambdaCjets}. For these measurements jets are reconstructed using Fastjet with the anti-$k_{\mathrm{T}}$ algorithm using a cone radius of $R=0.4$, $|\eta_{\mathrm{J}}| < 0.5$ and $7 < p_{\mathrm{T}}^{\mathrm{J}} < 15$ \gevc while charm hadrons are required to have $|y^{\mathrm{H}}| < 0.8$ and $3 < p_{\mathrm{T}}^{\mathrm{H}} < 15$ \gevc \cite{FastJet,AntiKtAlgo}. The variable used is the jet momentum due to the charm hadron along the direction of the jet axis:

$$z_{||} = \frac{\overline{p_{\mathrm{J}}} \, \cdot \, \overline{p_{\mathrm{H}}}}{\overline{p_{\mathrm{J}}} \, \cdot \, \overline{p_{\mathrm{J}}}}$$

with 1 and 0 as upper and lower limits, respectively. On one hand, when \zjet $\approx$ 1 the jet momentum is entirely carried by the charm hadron as it is the only constituent. On the other hand, \zjet $\approx$ 0 indicates an important amount of jet activity as the contribution to the jet momentum from the charm hadron is minimal. 


\begin{figure*}
\begin{center}
\includegraphics[width=0.49\textwidth]{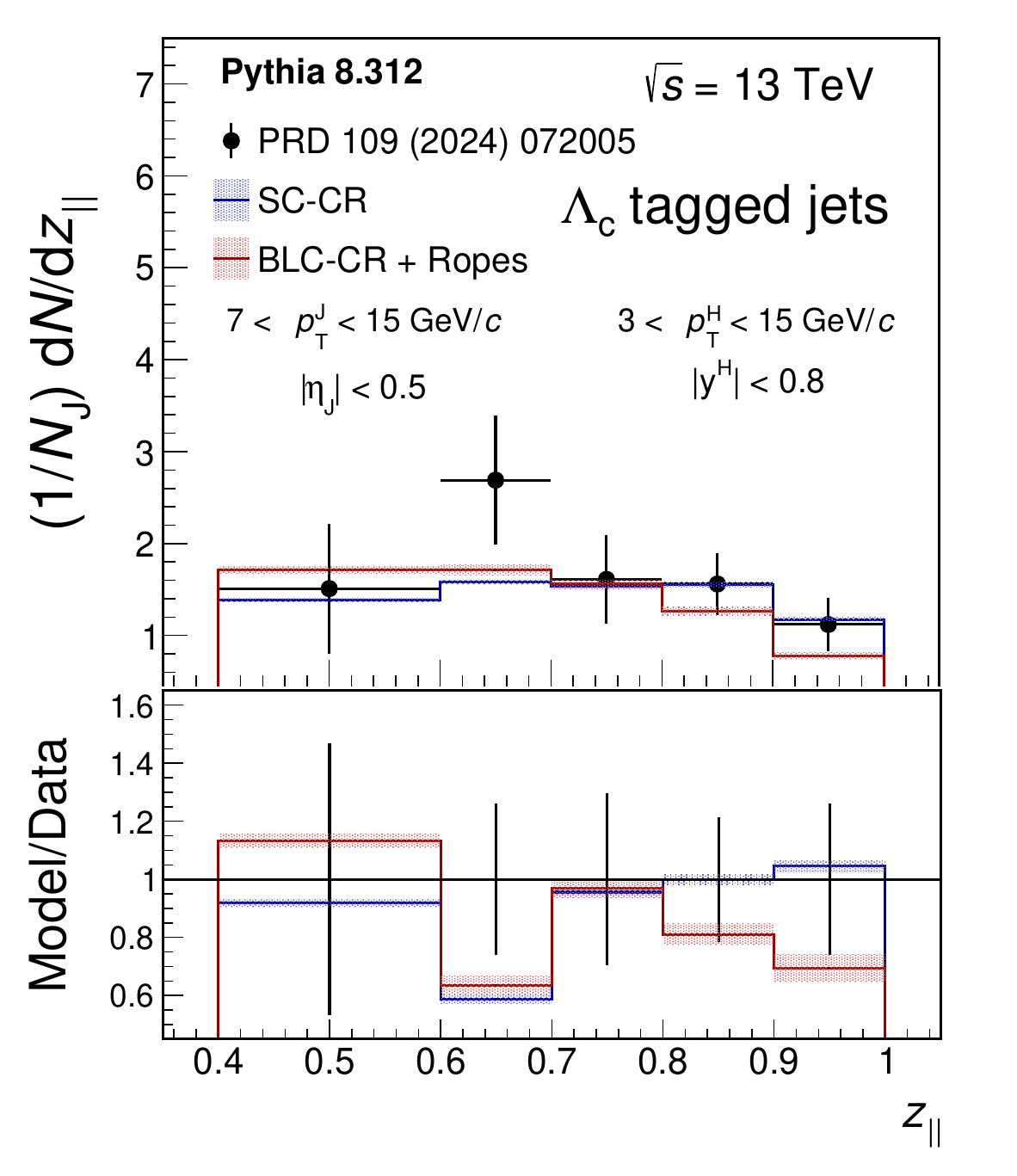}
\hspace{-0.4cm}
\includegraphics[width=0.49\textwidth]{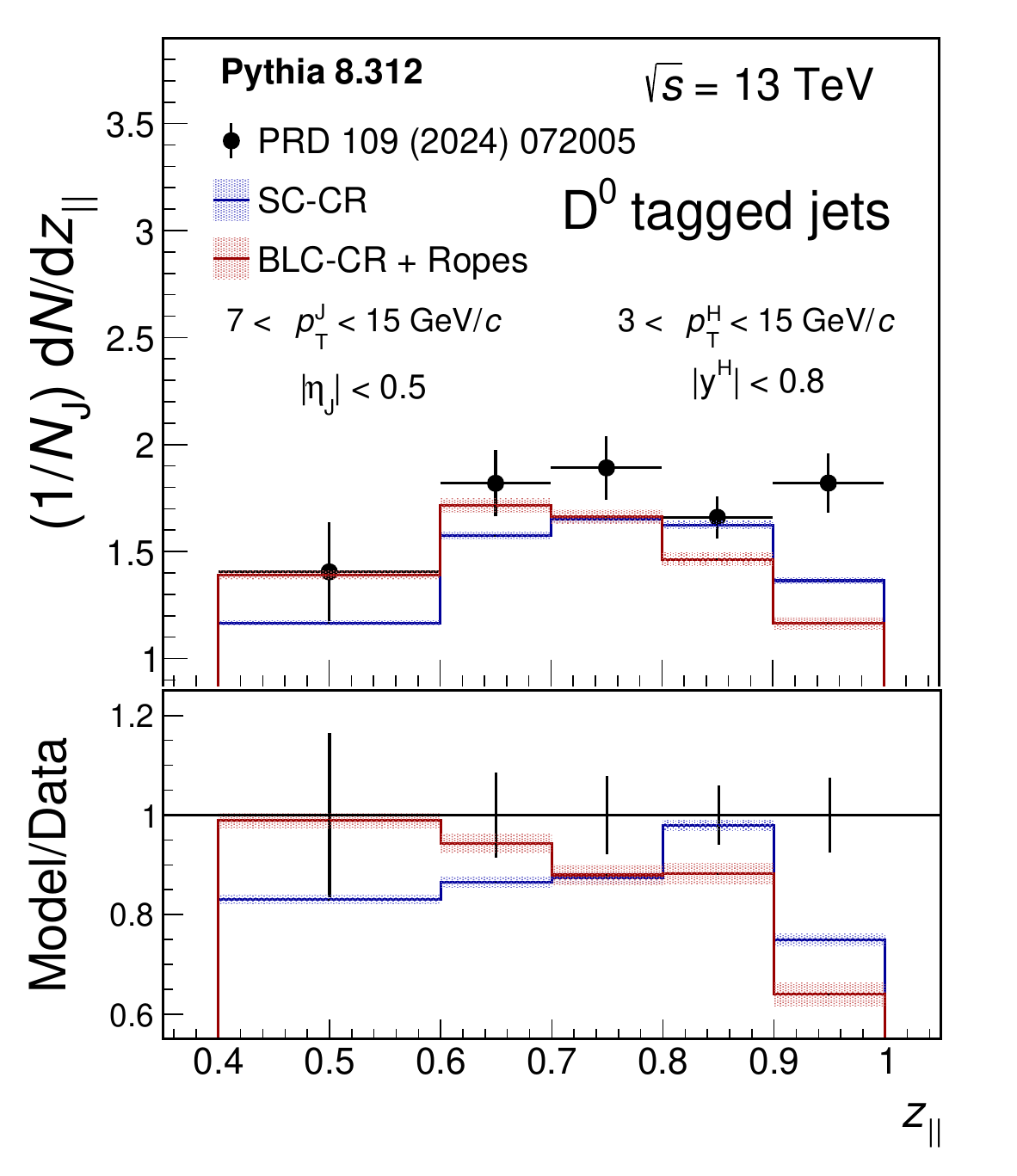}
\caption{Normalized \zjet distributions for $\Lambda_c^+$ (left) and D$^0$ (right) tagged jets. Black markers are the experimental data where the vertical bars are the quadratic sum of the statistical and systematic uncertainties while the horizontal bars indicate the bin width. Full lines represent the \sccr (blue) and \blc (red) predictions. Shaded areas around the lines are the statistical uncertainties from the simulation. Bottom panels show the model to data ratios}
\label{DvsSjets}
\end{center}
\end{figure*}

\begin{figure*}
\begin{center}
\includegraphics[width=0.49\textwidth]{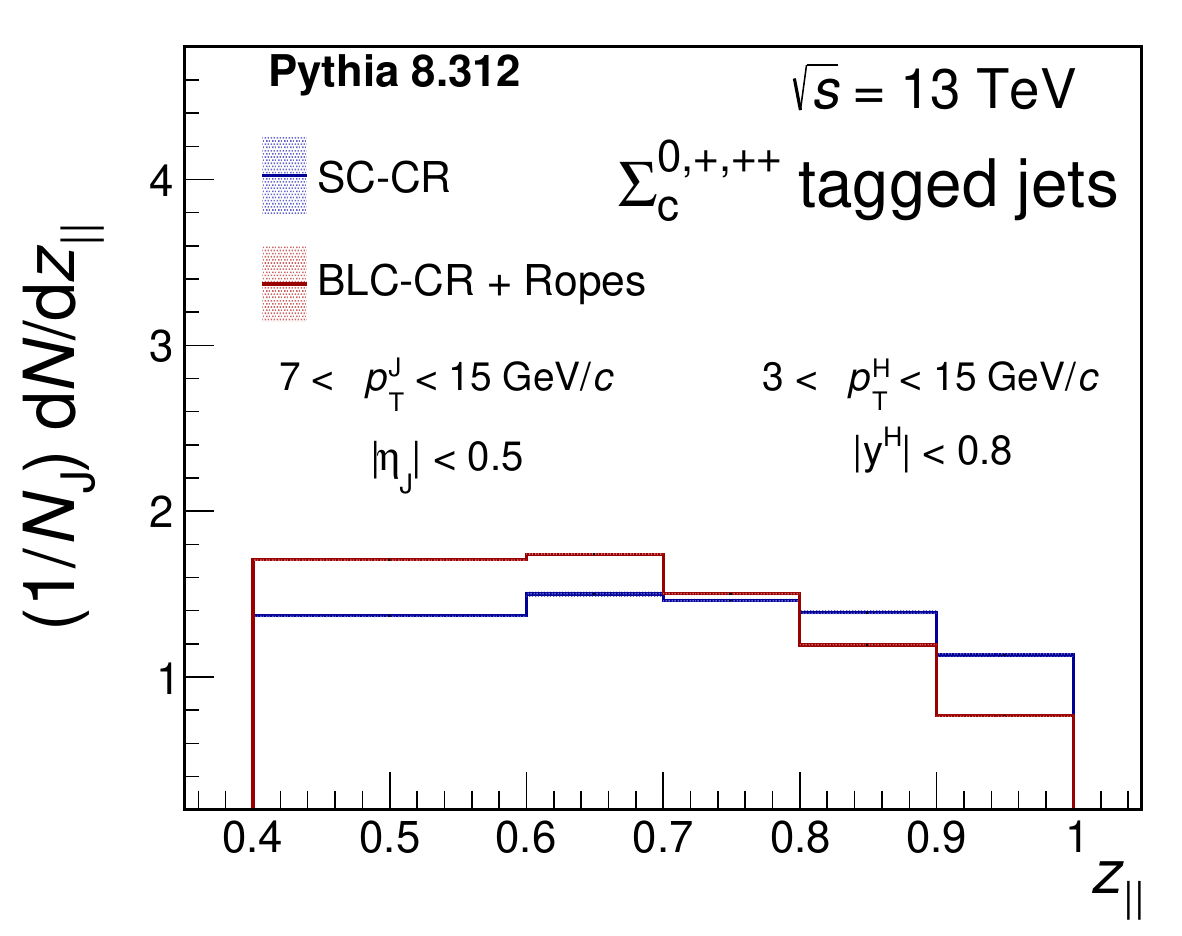}
\hspace{-0.4cm}
\includegraphics[width=0.49\textwidth]{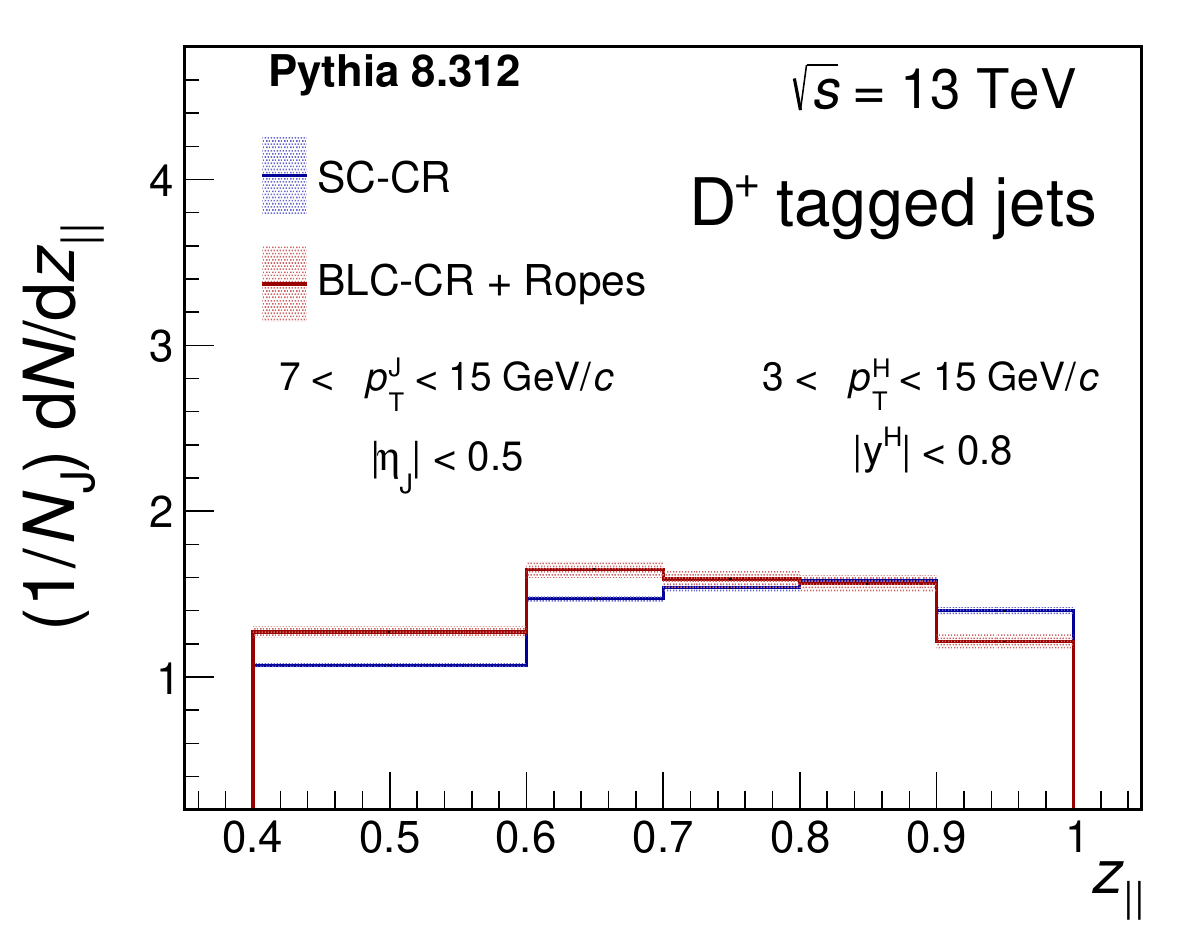}
\caption{Normalized \zjet distributions for $\Sigma_c^{0,+,++}$ (left) and D$^+$ (right) tagged jets. Full lines represent the \sccr (blue) and \blc (red) predictions. Shaded areas around the lines are the statistical uncertainties from the simulation.}
\label{PredictedJets}
\end{center}
\end{figure*}


Figure \ref{DvsSjets} shows the predictions for the $\Lambda_c^+$ and D$^0$ results provided by the \sccr and \blc models in blue and red lines, respectively. Black markers are the experimental data where the vertical bars are the quadratic sum of the statistical and systematic uncertainties while the horizontal bars indicate the bin width. Shaded areas around the full lines are the statistical uncertainties from the simulation. Bottom panels show the model to data ratios and the vertical bars in the unit line are the errors from the data. Both models have a very good performance reproducing the data with a small under estimation when \zjet $\approx$ 1 for the D$^0$ meson. Notice the lowest \zjet value is limited by the minimum \pth and \ptj from the kinematic selection. Figure \ref{PredictedJets} shows the predictions of how the $\Sigma_c^{0,+,++}$ and D$^+$ in jets distribution would look like with the same kinematical requirements as the $\Lambda_c^+$ and D$^0$. For D$^{*+}$ a very similar result is obtained as in the D$^{+}$ case.

\begin{figure*}
\begin{center}
\includegraphics[width=0.98\textwidth]{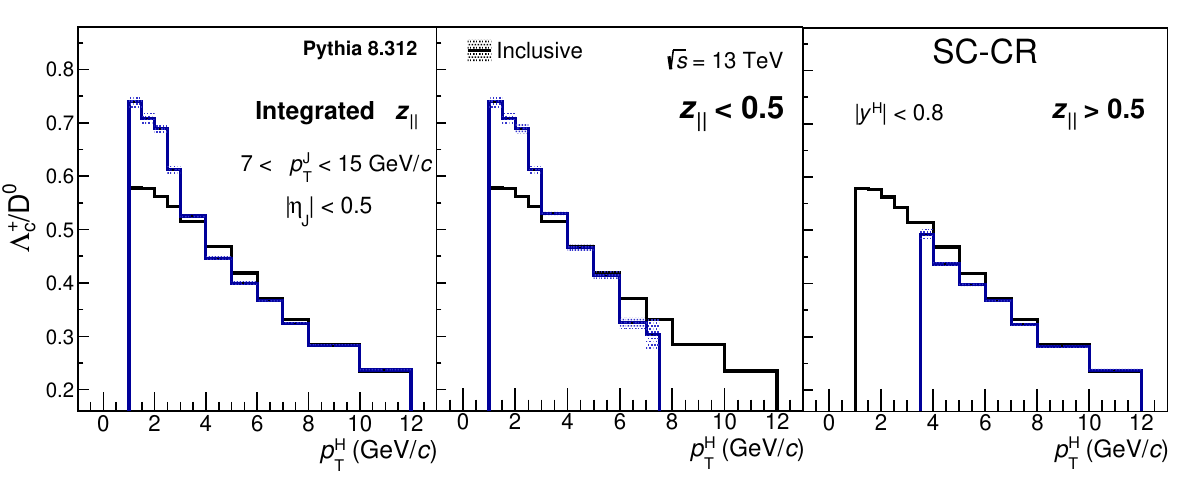}
\caption{\barionone ratios as a function of \pth when both hadrons are within jets according to \sccr model: \zjet integrated (left), \zjet $<$ 0.5 (middle) and \zjet $>$ 0.5 (right). Black line corresponds to the inclusive case. Shaded areas indicate the statistical uncertainties.}
\label{L2D0JetsSCCR}
\end{center}
\end{figure*}

\begin{figure*}
\begin{center}
\includegraphics[width=0.98\textwidth]{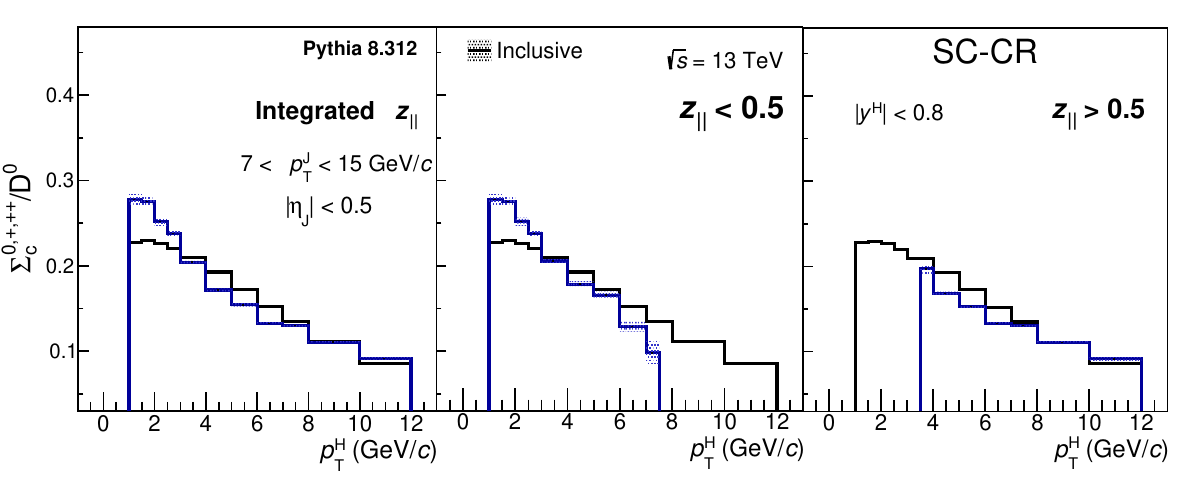}
\caption{\bariontwo ratios as a function of \pth when both hadrons are within jets according to \sccr model: \zjet integrated (left), \zjet $<$ 0.5 (middle) and \zjet $>$ 0.5 (right). Black line corresponds to the inclusive case. Shaded areas indicate the statistical uncertainties.}
\label{S2D0JetsSCCR}
\end{center}
\end{figure*}

\begin{figure*}
\begin{center}
\includegraphics[width=0.98\textwidth]{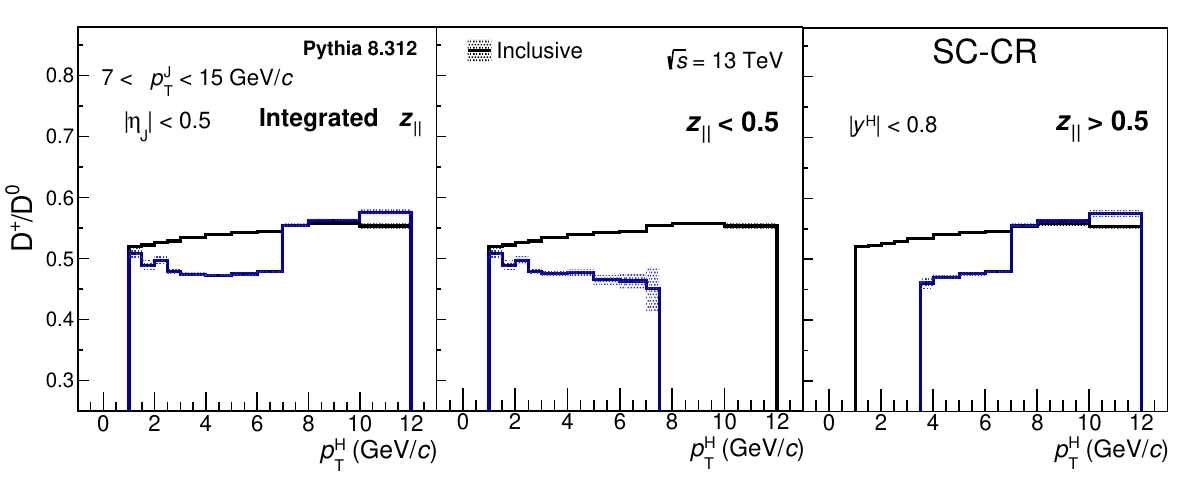}
\caption{\mesonone ratios as a function of \pth when both hadrons are within jets according to \sccr model: \zjet integrated (left), \zjet $<$ 0.5 (middle) and \zjet $>$ 0.5 (right). Black line corresponds to the inclusive case. Shaded areas indicate the statistical uncertainties.}
\label{Dp2D0JetsSCCR}
\end{center}
\end{figure*}

\begin{figure*}
\begin{center}
\includegraphics[width=0.98\textwidth]{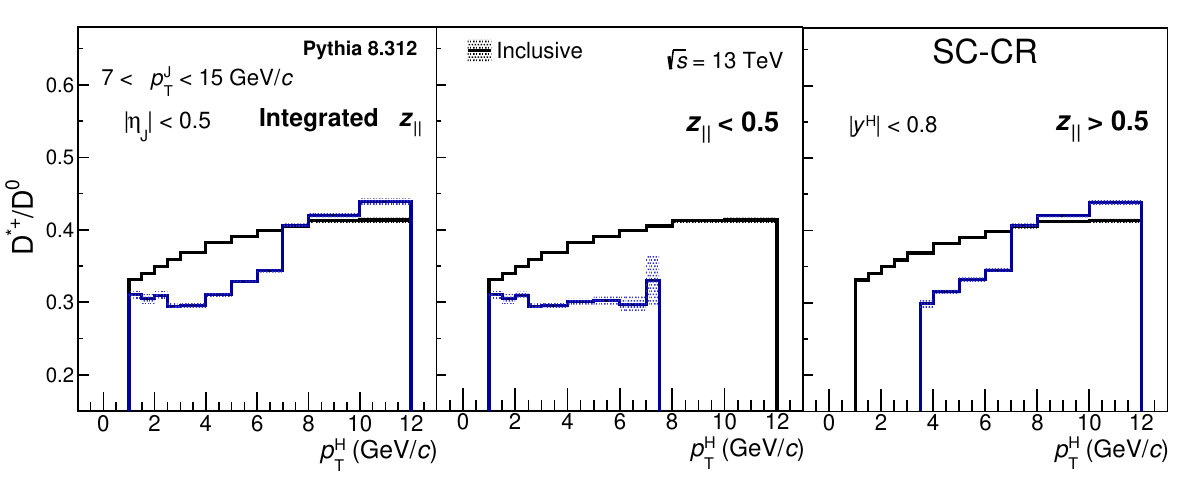}
\caption{\mesontwo ratios as a function of \pth when both hadrons are within jets according to \sccr model: \zjet integrated (left), \zjet $<$ 0.5 (middle) and \zjet $>$ 0.5 (right). Black line corresponds to the inclusive case. Shaded areas indicate the statistical uncertainties.}
\label{Ds2D0JetsSCCR}
\end{center}
\end{figure*}

The next step is to compute the \barionone, \bariontwo, \mesonone and \mesontwo ratios when both hadrons are found within jets, using the integrated \zjet range, and compare to the case when no jet finding algorithm is employed (inclusive events). The latter are those shown in figures \ref{DvsSbaryons} and \ref{DvsSmesons}, but with the slight modification of the rapidity coverage that is now increased, so $|y^{\mathrm{H}}| < 0.8$. This is motivated in order to have the same kinematical ranges for a better comparison. For the same reason, the transverse momentum of the charm hadron in the jet is extended down to \pt = 1 \gevc. According to \sccr and \blc models there is a clear difference among the ratios depending if the charm resonances are found within jets (left panels of figures \ref{L2D0JetsSCCR} to \ref{Ds2D0JetsSCCR}). For \mesonone and \mesontwo the effect of being jet constituents is mainly visible in the $1 < p_{\mathrm{T}}^{\mathrm{H}} < 7$ \gevc range while for \barionone and \bariontwo the impact is predominant at low \pt. This is a clear indication that the hadronisation of charm quarks is affected by local conditions. Indeed, components of the charm hadron states immerse in jets could undergo re-scattering with the accompanying particles so the fragmentation ratios are modified relative to the inclusive events. Similar results are extracted from the \blc model (plots located in the appendix).


Further refinement to these results can be achieved by splitting them into two categories: \zjet $<$ 0.5 and \zjet $>$ 0.5. Results for the lower (upper) \zjet range are located in the middle (right) panels of figures \ref{L2D0JetsSCCR} to \ref{Ds2D0JetsSCCR}. On one hand the maximum transverse momentum that can be reached in the \zjet $<$ 0.5 range is \pth = 7.5 \gevc, as it is limited by the highest \ptj value. On the other hand, for \zjet $>$ 0.5, the \pth = 3.5 \gevc boundary is provided by the lowest \ptj value. For \barionone and \bariontwo, the effect caused by being constituents of jets mostly impacts in the lower \zjet $<$ 0.5 range. This is expected as in the \zjet integrated case the difference with respect to the inclusive events is mainly in the low \pth sector. For \mesonone and \mesontwo, as the deviation from the inclusive ratio spams 2 $<$ \pth $<$ 7 \gevc, then both \zjet ranges are sensitive to this effect.

\section{Relative transverse activity}

As shown in figure \ref{DvsSmult} there is a clear multiplicity dependence of the \barionone ratio. However, by choosing high multiplicity events the data set is intrinsically biased as hard processes dominate the sample. In this sense event shape observables are very useful to reduce the selection bias by diminishing the susceptibility to hard processes. Furthermore, as event shape observables are infrared and collinear safe they highly valuable tools in QCD studies \cite{EventShapes}. 


A pp collision can be split into a hard scattering component and the underlying event (MPI, beam remnants, etc.) that comprises the soft interactions. A well known method to perform this separation is by taking the leading particle (the one with the largest \pt) as a reference to divide the azimuthal plane into three regions. Setting $\Delta \phi$ as the azimuthal angular difference between the leading particle and any other, it is possible to define the towards ($|\Delta\phi|$ $<$ $\pi/3$), away ($|\Delta\phi|$ $>$ $2\pi /3$) and transverse ( $\pi /3$ $<$ $|\Delta\phi|$ $<$ $2\pi /3$) regions. The latter being the less sensitive to the hard interaction.

The relative transverse activity classifier ($R_{\mathrm{T}}$) is an example of an event shape observable and is defined as \cite{Rt}:

$$ R_{\mathrm{T}} = \frac{N_{\mathrm{ch}}^{\mathrm{T}}}{\langle N_{\mathrm{ch}}^{\mathrm{T}} \rangle} $$

where $N_{\mathrm{ch}}^{\mathrm{T}}$ is the number of charged particles measured in the transverse region. So, the \rt is a normalized indicator of the underlying event (UE) activity and splits events into two categories: higher than average UE and lower than average UE. 

For this study the selection cuts follow those applied by the ALICE experiment in a recent publication where the relative transverse activity classifier is also used \cite{RtALICE}. The requirement on the leading particle is fulfilled if $8 < p_{\mathrm{T}}^{\mathrm{L}} < 15$ \gevc, for $N_{\mathrm{ch}}^{\mathrm{T}}$ only primary charged particles are used with \pt $>$ 0.5 \gevc and $|\eta| < 0.8$. Decay products from charm hadrons are not considered for the number of particles in the transverse region.

\begin{figure}[H]
\begin{center}
\includegraphics[width=0.49\textwidth]{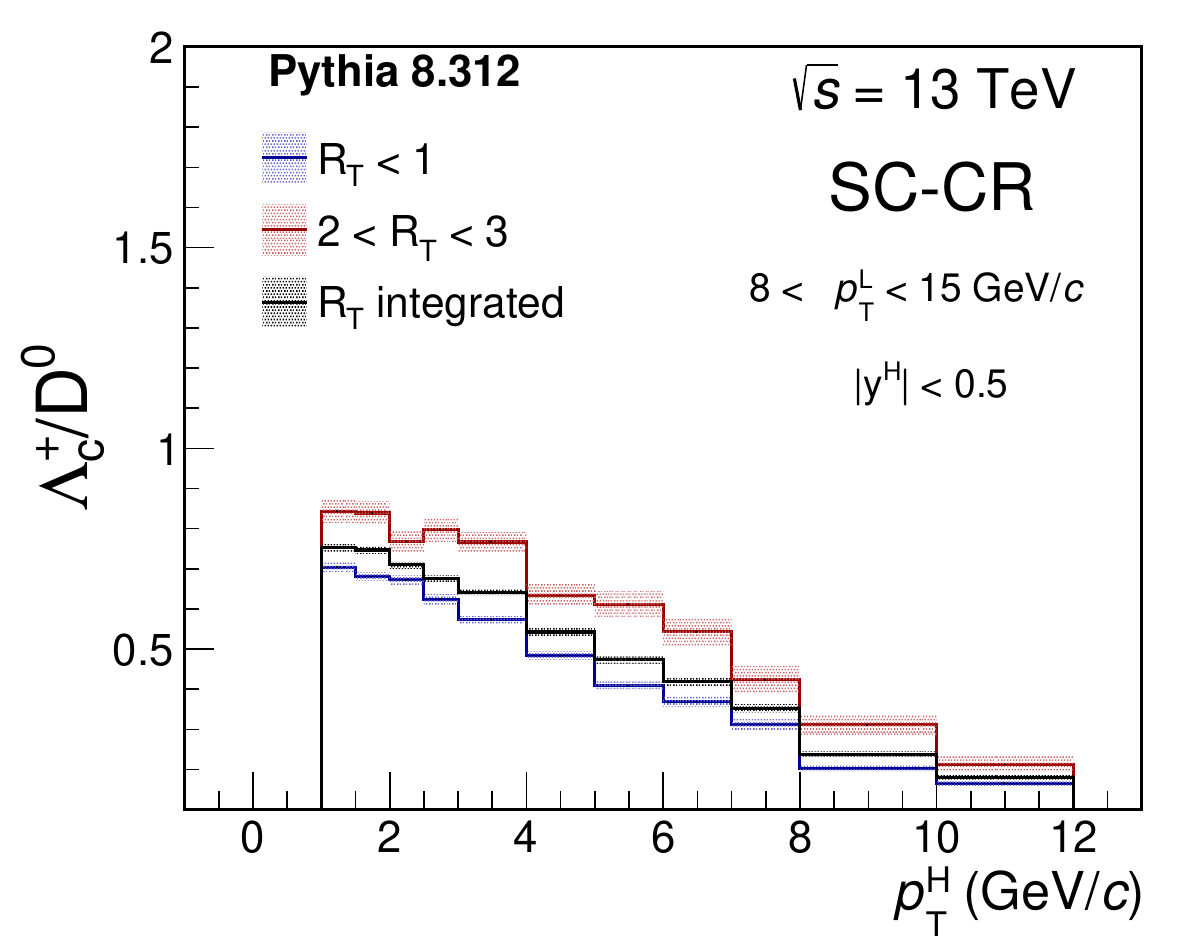}
\hspace{-0.4cm}
\includegraphics[width=0.49\textwidth]{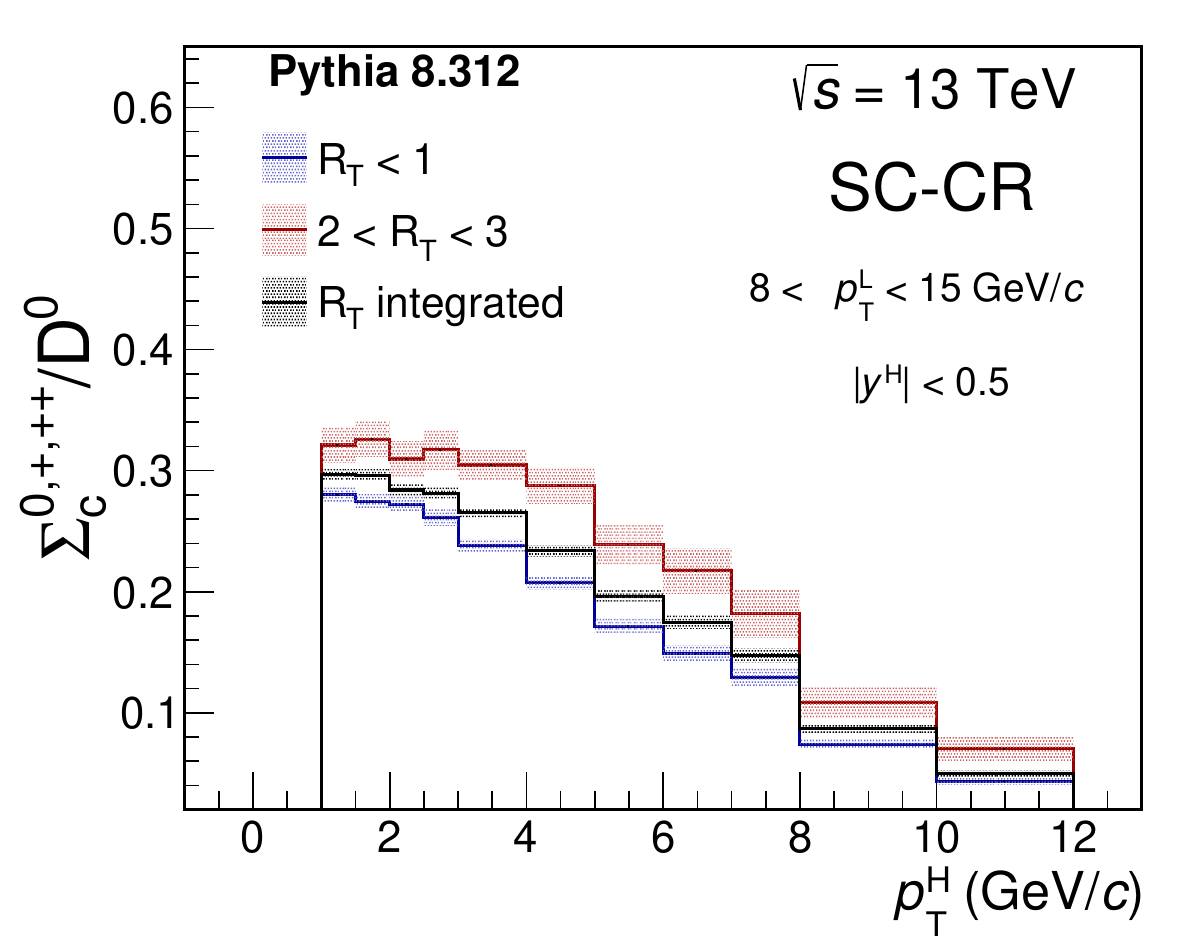}
\caption{\barionone (left) and \bariontwo (right) ratios as a function of \pth for two different relative transverse activity ranges according to the \sccr model. Black line corresponds to the \rt integrated, blue is for \rt $<$ 1 and red for 2 $<$ \rt $<$ 3. Shaded areas around the lines are the statistical uncertainties from the simulation.}
\label{RtBaryonsSCCR}
\end{center}
\end{figure}

\begin{figure}[H]
\begin{center}
\includegraphics[width=0.49\textwidth]{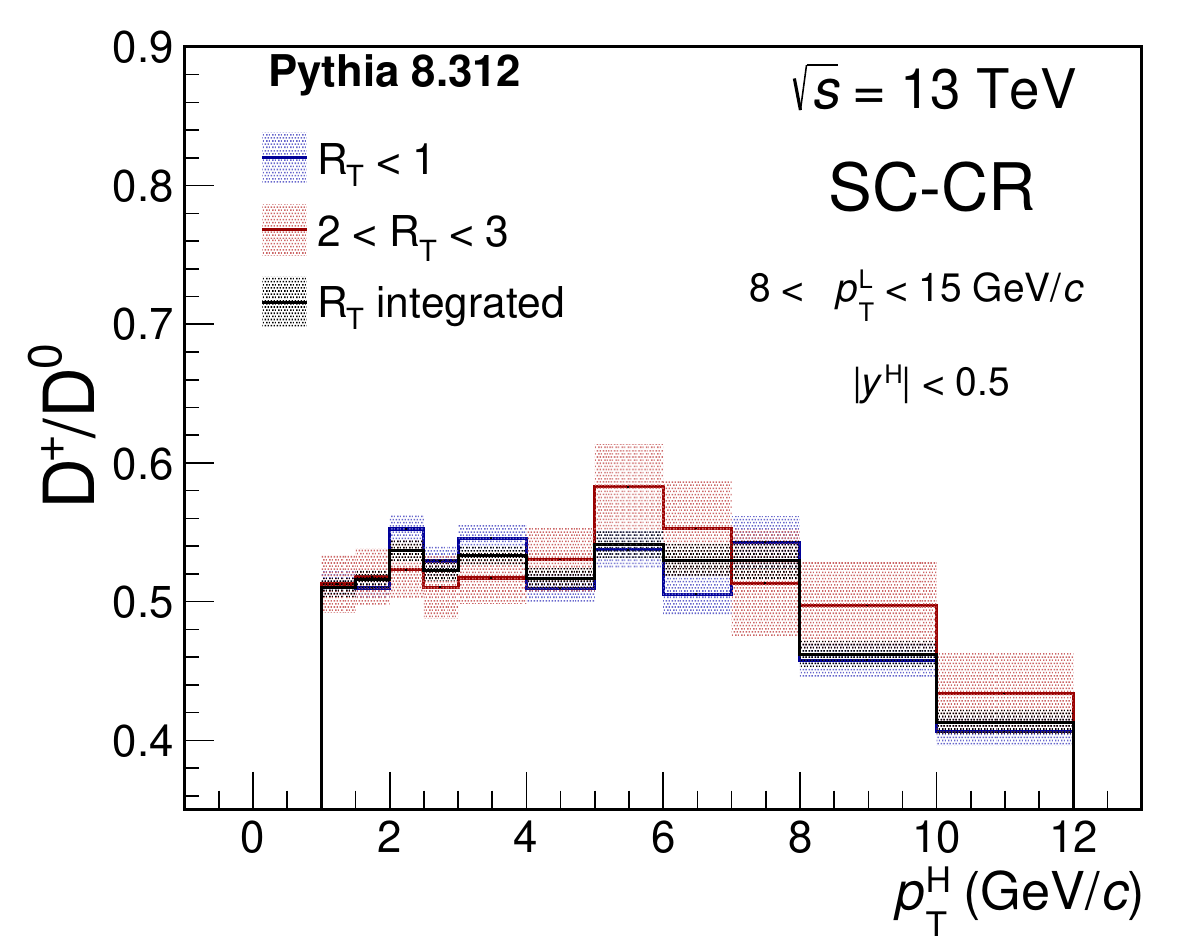}
\hspace{-0.4cm}
\includegraphics[width=0.49\textwidth]{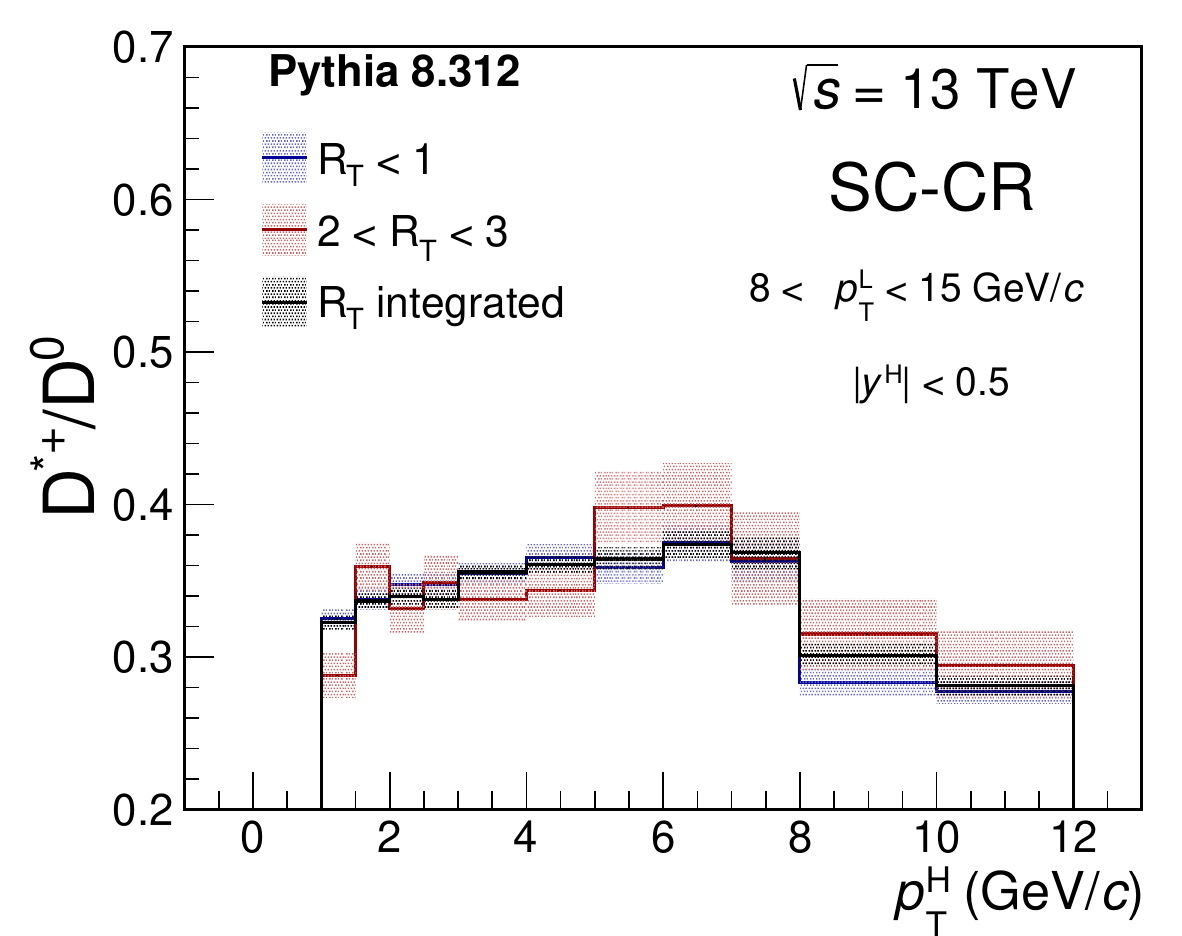}
\caption{\mesonone (left) and \mesontwo (right) ratios as a function of \pth for two different relative transverse activity ranges according to the \sccr model. Black line corresponds to the \rt integrated, blue is for \rt $<$ 1 and red for 2 $<$ \rt $<$ 3. Shaded areas around the lines are the statistical uncertainties from the simulation.}
\label{RtMesonsSCCR}
\end{center}
\end{figure}

Figure \ref{RtBaryonsSCCR} shows the \barionone and \bariontwo ratios for two different ranges of \rt according to the \sccr model. The black line is the integrated one, while the blue and red lines correspond to \rt $<$ 1 and 2 $<$ \rt $<$ 3, respectively. These lines indicate a low (blue) and large (red) UE activity, that is in turn correlated to the number of MPI \cite{RtMPI}. These baryon to meson ratios have an important dependence on the relative transverse activity, confirming the result shown in figure \ref{DvsSmult} but reducing the bias due to the multiplicity selection. For the \mesonone and \mesontwo in figure \ref{RtMesonsSCCR}, the \sccr model indicates that the ratios are completely independent of the UE activity. Therefore \rt $<$ 1 and 2 $<$ \rt $<$ 3 overlap, within uncertainties, with the integrated distribution. Similar results are extracted from the \blc model in the appendix.

For baryons these results can be understood by the fact that the \rt increases with the UE activity and in turn with the number of MPI, so color lines increase the reconnection with the different partons emerging from the augmented set of MPI. As a consequence the string tension is reduced, leading to the string breaking that enhances baryon production over mesons. For mesons it points to an independent fragmentation process: charm quark hadronisation for mesons is irrespective of the dense environment generated by the different MPI. All in all, global event conditions as the increase on the number of MPI, has not the same effect on charm baryon and meson hadronisation.

\section{Conclusion}

Recent experimental data indicate that the universality of the fragmentation functions for heavy hadrons is no longer valid. As a consequence a new set of models and Pythia 8 tunes, not relying on the fragmentation functions extracted from e$^-$p and e$^-$e$^+$ collisions, had to be developed. In this work the \blc and \sccr models are used to study prompt charm hadron ratios in different event conditions. In particular they have been confronted to recent measurements of \barionone, \bariontwo, \mesonone and \mesontwo ratios from the ALICE experiment in pp collisions at \s = 13 TeV. In general, both models have shown a good performance to reproduce the data. When the hadrons are found within jets, these baryon to meson and meson to meson ratios show deviations relative to the inclusive distributions. Such behavior could be the aftermath of a re-scattering with the accompanying particles inside the jet. A study focusing in the global event condition was also carried out. To this end the prompt charm hadron ratios have been computed in two different \rt ranges. Baryon to meson ratios indicate a clear dependence on the UE activity, while meson to meson quotients have no variations at all. The explanation is that, while charm baryon hadronization is modified by an increase in the numer of MPI, for mesons this points to an independent fragmentation process.


\section{Acknowledgments}

This work has employed an important amount of computing resources from the ACARUS at the Universidad de Sonora, without this facility it would have been impossible to develop this study.

\section*{Appendix}


\begin{figure}[H]
\begin{center}
\includegraphics[width=0.98\textwidth]{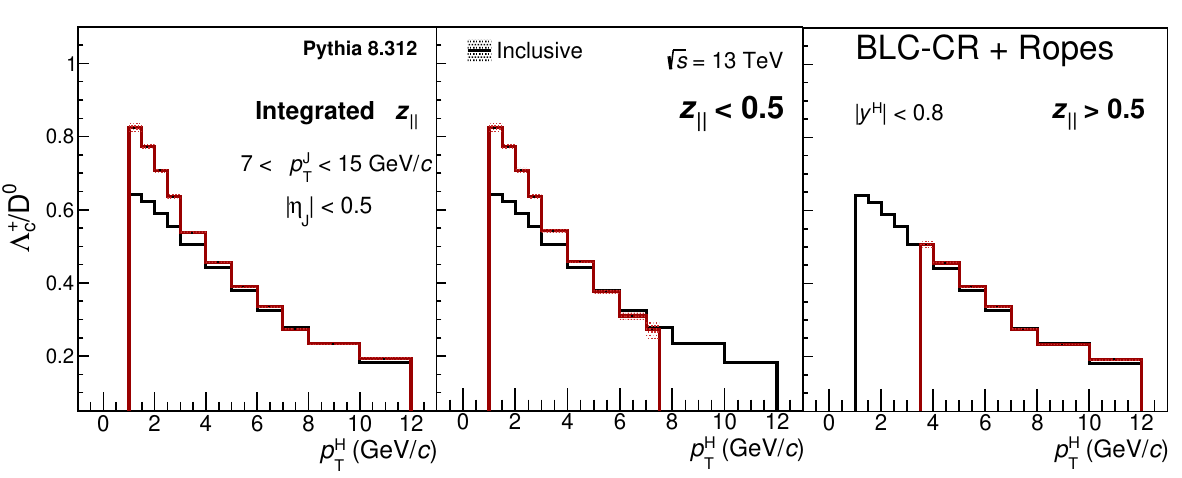}
\caption{\barionone ratios as a function of \pth when both hadrons are within jets according to \blc model: \zjet integrated (left), \zjet $<$ 0.5 (middle) and \zjet $>$ 0.5 (right). Black line corresponds to the inclusive case. Shaded areas indicate the statistical uncertainties.}
\label{L2D0JetsNewModel}
\end{center}
\end{figure}

\begin{figure}[H]
\begin{center}
\includegraphics[width=0.98\textwidth]{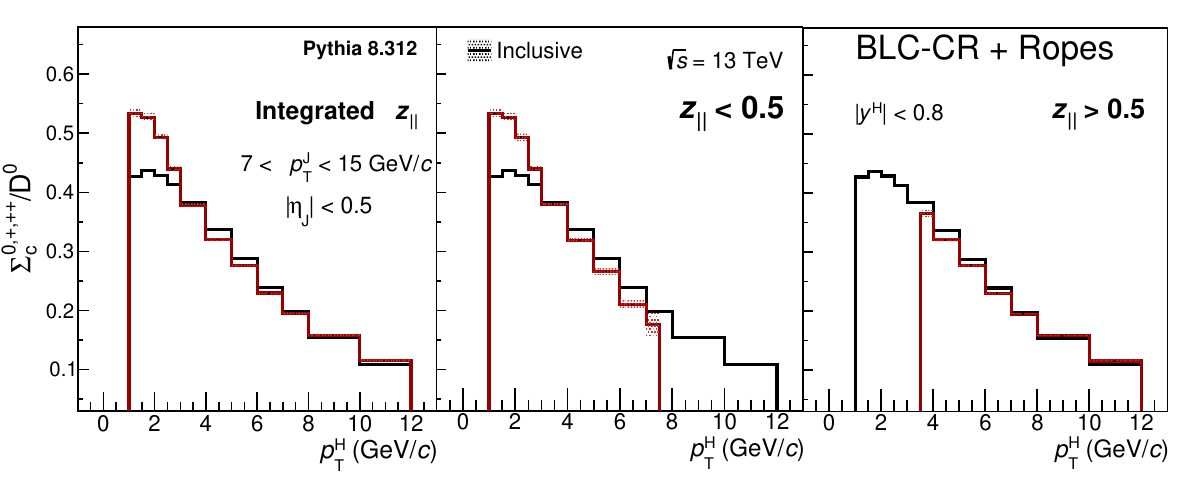}
\caption{\bariontwo ratios as a function of \pth when both hadrons are within jets according to \blc model: \zjet integrated (left), \zjet $<$ 0.5 (middle) and \zjet $>$ 0.5 (right). Black line corresponds to the inclusive case. Shaded areas indicate the statistical uncertainties.}
\label{S2D0JetsNewModel}
\end{center}
\end{figure}

\begin{figure}[H]
\begin{center}
\includegraphics[width=0.98\textwidth]{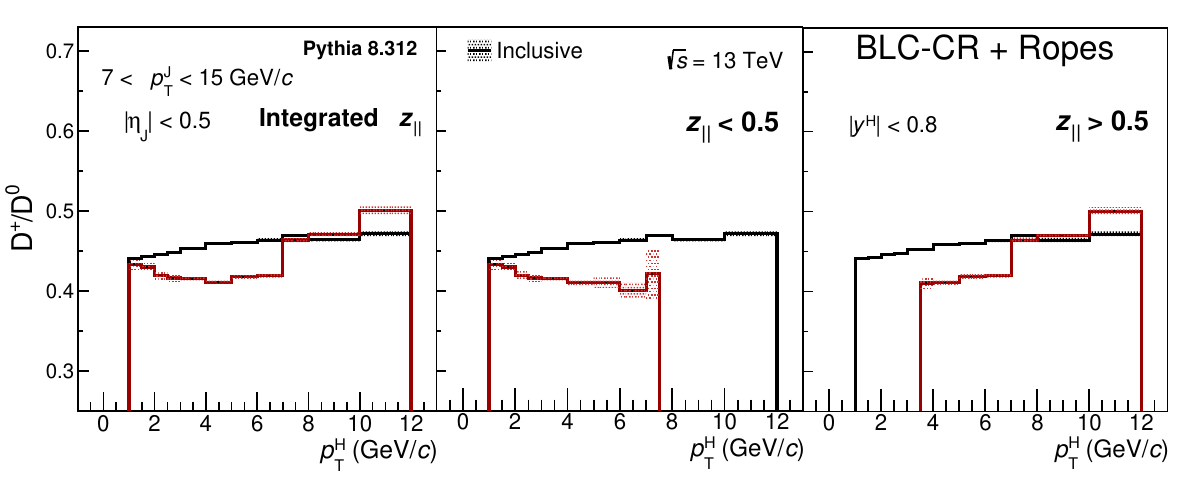}
\caption{\mesonone ratios as a function of \pth when both hadrons are within jets according to \blc model: \zjet integrated (left), \zjet $<$ 0.5 (middle) and \zjet $>$ 0.5 (right). Black line corresponds to the inclusive case. Shaded areas indicate the statistical uncertainties.}
\label{Dp2D0JetsNewModel}
\end{center}
\end{figure}

\begin{figure}[H]
\begin{center}
\includegraphics[width=0.98\textwidth]{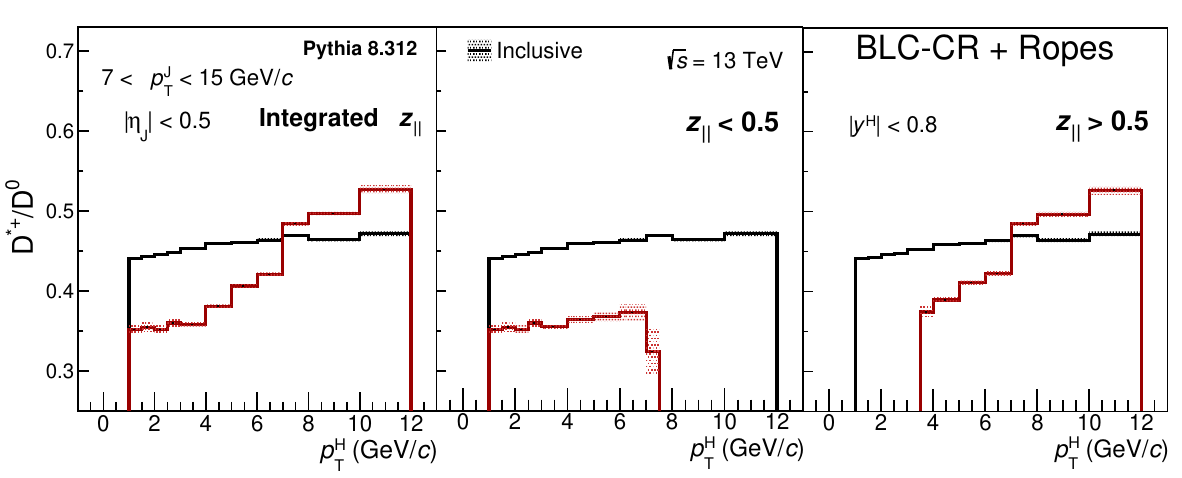}
\caption{\mesontwo ratios as a function of \pth when both hadrons are within jets according to \blc model: \zjet integrated (left), \zjet $<$ 0.5 (middle) and \zjet $>$ 0.5 (right). Black line corresponds to the inclusive case. Shaded areas indicate the statistical uncertainties.}
\label{Ds2D0JetsNewModel}
\end{center}
\end{figure}

\begin{figure}[H]
\begin{center}
\includegraphics[width=0.49\textwidth]{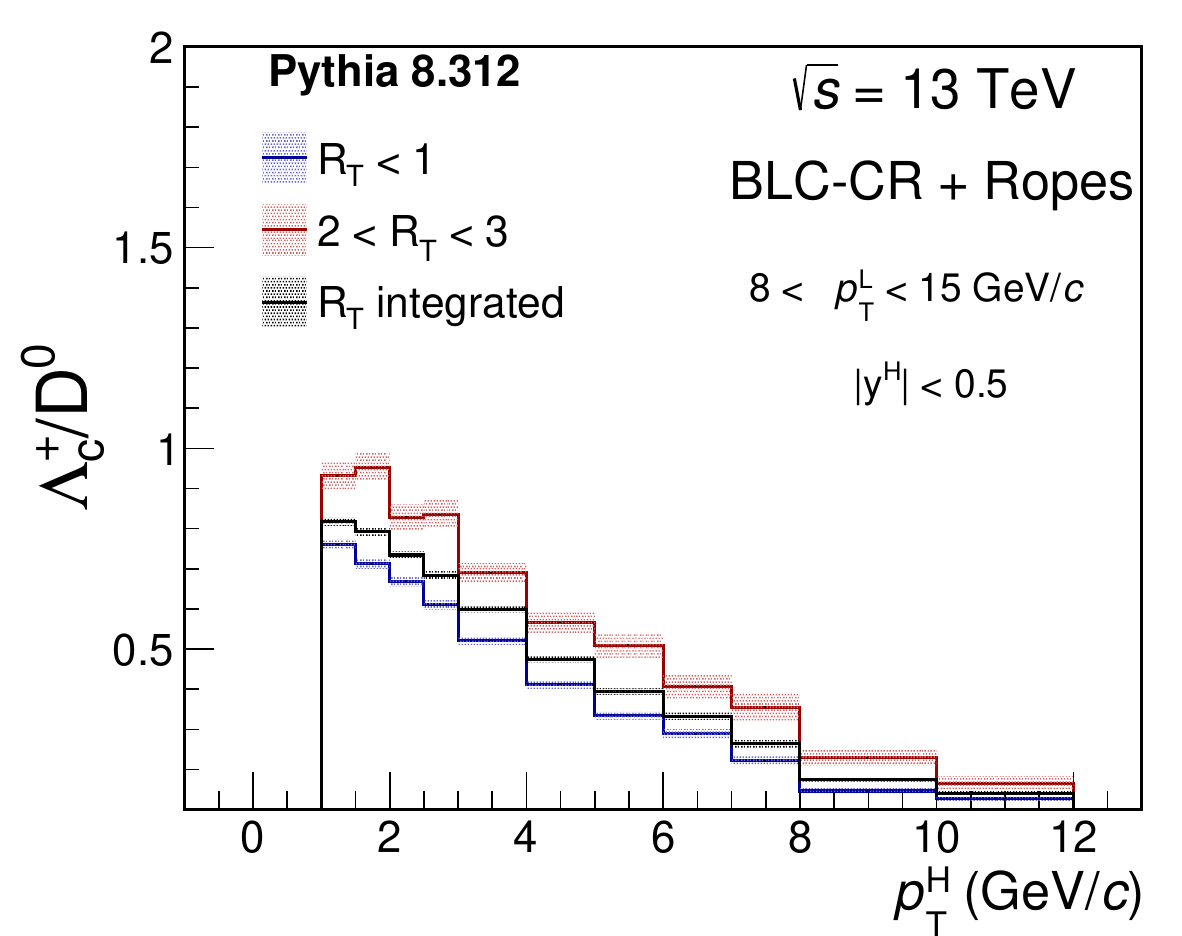}
\hspace{-0.4cm}
\includegraphics[width=0.49\textwidth]{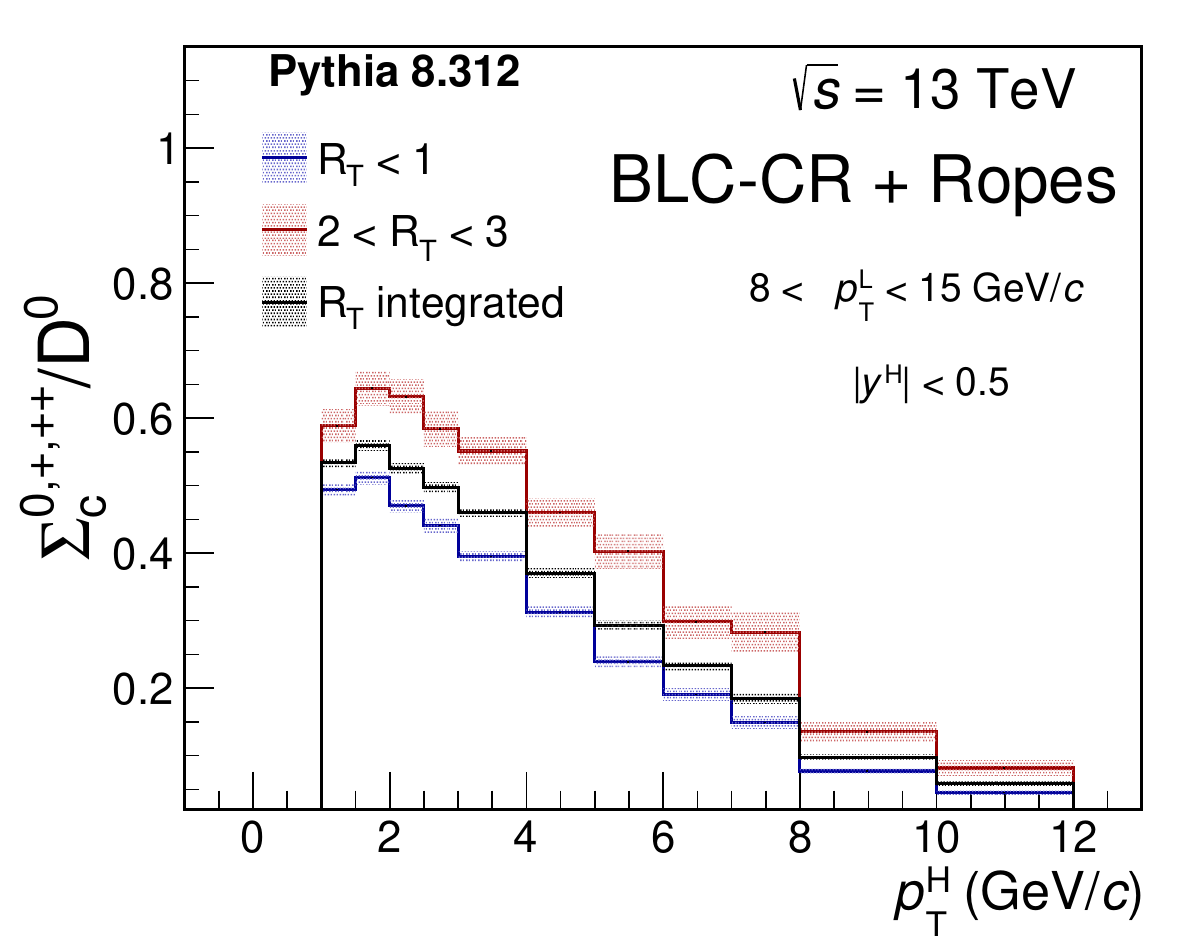}
\caption{\barionone (left) and \bariontwo (right) ratios as a function of \pth for two different relative transverse activity ranges according to the \blc model. Black line corresponds to the \rt integrated, blue is for \rt $<$ 1 and red for 2 $<$ \rt $<$ 3. Shaded areas around the lines are the statistical uncertainties from the simulation.}
\label{RtBaryonsNewModel}
\end{center}
\end{figure}

\begin{figure}[H]
\begin{center}
\includegraphics[width=0.49\textwidth]{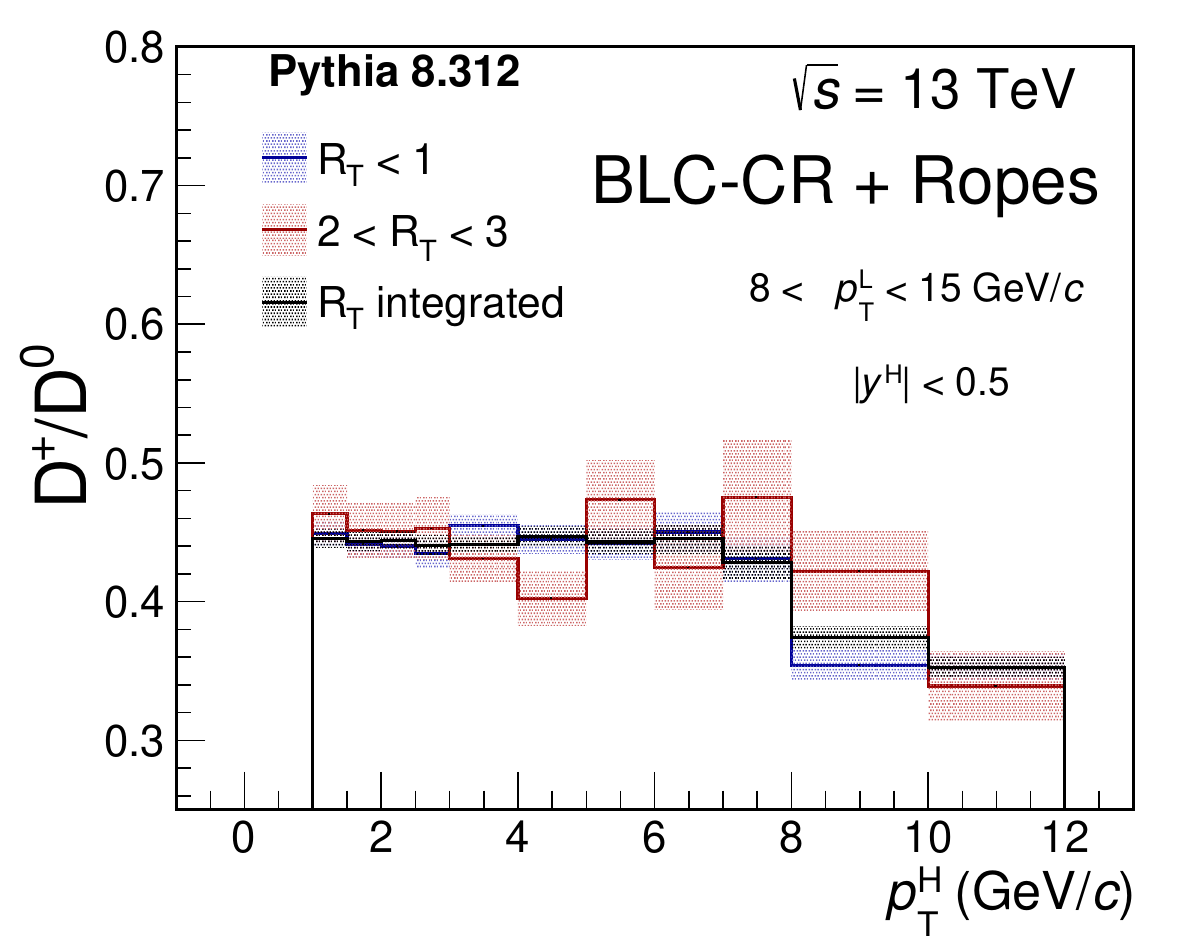}
\hspace{-0.4cm}
\includegraphics[width=0.49\textwidth]{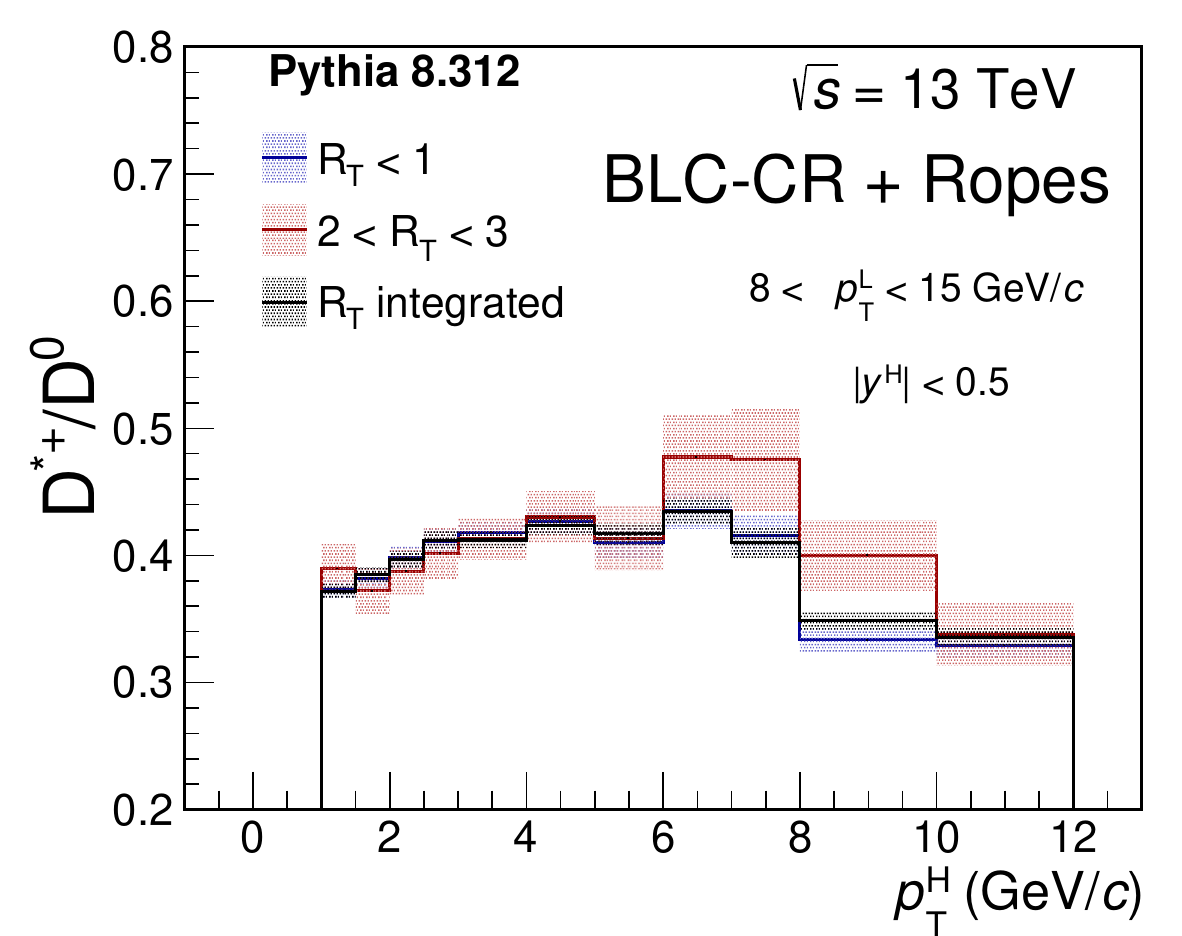}
\caption{\mesonone (left) and \mesontwo (right) ratios as a function of \pth for two different relative transverse activity ranges according to the \blc model. Black line corresponds to the \rt integrated, blue is for \rt $<$ 1 and red for 2 $<$ \rt $<$ 3. Shaded areas around the lines are the statistical uncertainties from the simulation.}
\label{RtMesonsNewModel}
\end{center}
\end{figure}


\bibliography{mybibfile}

\end{document}